\documentclass{article}

\PassOptionsToPackage{numbers}{natbib}
\usepackage[main, final]{neurips_2026}

\usepackage[utf8]{inputenc} 
\usepackage[T1]{fontenc}    
\usepackage{hyperref}       
\usepackage{url}            
\usepackage{booktabs}       
\usepackage{amsfonts}       
\usepackage{nicefrac}       
\usepackage{microtype}      
\usepackage{xcolor}         
\usepackage{wrapfig}
\usepackage{graphicx}
\usepackage{algorithm}
\usepackage{algorithmic}  
\usepackage{float}
\usepackage{enumitem}
\usepackage{ulem}
\usepackage{makecell}
\usepackage{graphicx}
\usepackage{array}
\usepackage{tabularx}
\usepackage{subcaption}
\usepackage{caption}

\usepackage{amsmath}
\title{CodeMimicry: Exploiting Safety Generalization Lag in Large Language Models via Structured Code Completion}

\author{%
Zhen Liang$^{1,2}$ \quad Hai Huang$^{1,2}$\thanks{Corresponding author. Email: haihuang1005@gmail.com} \quad Wentao Chen$^{3}$ \\ $^{1}$School of Computer Science and Technology, Zhejiang Sci-Tech University \\ $^{2}$Zhejiang Key Laboratory of Digital Fashion and Data Governance, Zhejiang Sci-Tech University \\Hangzhou 310018, China \\ $^{3}$China Academy of Information and Communications Technology  \\ \texttt{\{liangzhen741, haihuang1005\}@gmail.com} \quad \texttt{chenwentao@caict.ac.cn}
  }

\begin{document}

\maketitle

\begin{abstract}
Large language models have achieved remarkable capabilities across diverse domains, yet their safety alignment remains vulnerable to jailbreak attacks. In this work, we identify a previously underexplored failure mode—safety generalization lag—where alignment trained predominantly on natural language fails to transfer to the code domain. We show that this lag induces a code-completion blind spot, allowing malicious intent embedded within syntactically valid code to evade safety mechanisms. To exploit this vulnerability, we propose CodeMimicry, a fully automated black-box jailbreak framework that generates structured, object-oriented code prompts to induce harmful outputs via code completion. Experiments on 8 state-of-the-art commercial LLMs demonstrate that CodeMimicry achieves a 96.25\% attack success rate with 1.51 queries on average, significantly outperforming both template-based and optimization-based baselines. Beyond empirical performance, we provide a mechanistic analysis of code-based jailbreaks through latent space representations, including projection onto refusal-related directions and activation steering. This analysis offers an explanation of how CodeMimicry bypasses safety mechanisms in code-related domains. Our findings reveal a weakness in current safety alignment and highlight the need for robust alignments in structured domains such as code. Our code is on \url{https://github.com/lzzzr123/CodeMimicry}.
\end{abstract}

\section{Introduction}
Large language models (LLMs) have demonstrated unprecedented capabilities in Natural Language Processing \cite{nlp, nlp2}, information retrieval \cite{research}, and code tasks \cite{code1, code2, code3}, driving their integration into critical applications. Therefore, ensuring these models adhere to security guidelines has become crucial. Techniques such as reinforcement learning based on human feedback (RLHF) \cite{rlhf} and supervised fine-tuning \cite{sft} are widely used to align models with human values, thus building guardrails against malicious abuse. However, LLMs remain vulnerable to sophisticated jailbreak attacks, in which carefully crafted malicious prompts can induce the model to produce harmful or unintended outputs.
\par Jailbreak attacks against LLMs are generally divided into two categories: (1) White-box attacks, such as GCG \cite{gcg}, which use gradient-based optimization to craft adversarial suffixes. However, they require access to model parameters and gradients, making them impractical for closed-source LLMs. (2) Black-box attacks, including template-based and automated methods. Template-based approaches such as CodeChameleon \cite{codechameleon} and FlipAttack \cite{flipattack} rely on manually designed prompts to bypass safety mechanisms, but are often mitigated by simple rule-based or pattern-based defenses due to their fixed structures. Automated methods such as PAIR \cite{pair}, GPTfuzzer \cite{gptfuzzer}, and AutoDan-Turbo \cite{autodanturbo} generate jailbreak prompts via model feedback, removing the need for manual design. However, they typically lack principled, domain-specific strategies (e.g., program structure or code completion), relying instead on heuristic exploration in natural language space, leading to limited effectiveness and efficiency.

\par A critical vulnerability arises from the rapid advancement of large language models: \emph{mismatched generalization} \cite{mg}, wherein models' broad pretraining capabilities substantially outpace the narrower scope of their safety alignment. While modern general LLMs have achieved near-expert proficiency in code generation and mathematical reasoning \cite{openai, gemini3}, their safety alignment is still predominantly trained on natural language data. Motivated by this observation, we hypothesize the existence of a \textbf{Safety Generalization Lag} in the code domain: \textit{as LLMs’ capabilities in the code domain continue to advance, their safety alignment fails to generalize effectively to this rigid, syntactic subspace.} 
This lag creates a \textbf{code-completion blind spot}, where safety-aligned representations fail to generalize to the structured code space. As refusal behaviors are primarily learned from natural language, they transfer poorly to code generation, allowing malicious intent embedded in syntactically valid code to bypass safeguards.

In this paper, to address these limitations and support our hypothesis, we propose CodeMimicry, an efficient and fully automated jailbreak framework that eliminates manual strategy design by automatically generating jailbreak prompts in which malicious intent is embedded within structured code.
Extensive experiments on 8 black-box commercial LLMs demonstrate the superiority of CodeMimicry. Notably, it achieves a 90\% or higher attack success rate on each model with a very low average of queries. 
To better understand the behavioral differences between successful and failed code-based jailbreaks, we further conduct an in-depth internal analysis of model representations and refusal behaviors.
These results suggest deficiencies in code-domain safety alignment and indicate that robust code-domain safety guardrails have not been sufficiently established within these LLMs. Our contributions are three-fold:
\begin{itemize}[leftmargin=0.5cm]
    \item \textbf{Code-domain Jailbreaking Utilizing Safety Generalization Lag:} 
    We identify a previously underexplored vulnerability in LLMs—safety generalization lag—where alignment does not reliably transfer to the code domain. Leveraging this insight, we propose CodeMimicry, an automated jailbreak framework that embeds malicious intent into structured object-oriented code, shifting models from safety-aligned dialogue to compliance-driven code completion.
    \item \textbf{Effectiveness with Minimal Query Budget:} 
    Across 8 state-of-the-art LLMs, CodeMimicry achieves a 96.25\% attack success rate with 1.51 queries on average on AdvBench, and 96.07\% ASR with 1.46 queries on HarmBench, substantially outperforming both template-based and optimization-based baselines.
    \item \textbf{Mechanistic Insight via Latent Space Analysis:} We conduct a mechanistic analysis of code-based jailbreaks in the latent representation space, including projection onto refusal-related directions and activation steering. The results show that successful jailbreaks remain close to benign code representations while avoiding refusal-associated features, providing an explanation of how CodeMimicry bypasses safety mechanisms in code domains.

\end{itemize}

\section{Related Work}

\textbf{In the white-box setting}, methods such as GCG \cite{gcg} and AutoDAN \cite{autodan} leverage gradient information to optimize discrete adversarial tokens. While effective, these approaches require direct access to model parameters and are therefore incompatible with commercial black-box APIs.\par
\textbf{In the black-box setting}, a substantial body of work employs diverse obfuscation and transformation strategies to conceal malicious objectives and thereby achieve jailbreak attacks.
FlipAttack \cite{flipattack}, CipherChat \cite{cipherchat}, ArtPrompt \cite{artprompt}, and DRA \cite{dra} obfuscate harmful instructions using Base64 or ASCII art. CodeChameleon \cite{codechameleon} and CodeAttack \cite{codeattack} embed malicious intent within code-like structures, while MathPrompt \cite{mathprompt} and EquaCode \cite{equacode} transform instructions into mathematical problem formulations.
Despite their effectiveness, these methods depend on manually crafted prompts to bypass the built-in safety mechanisms of LLMs. Their fixed and easily recognizable structures, however, make them vulnerable to simple rule- or pattern-based defenses.
Another line of work formulates jailbreaking as an iterative prompt optimization problem. Evolutionary approaches (e.g., GPTFuzzer \cite{gptfuzzer}) and attacker-LLM-based frameworks, including PAIR \cite{pair}, TAP \cite{tap}, GAP \cite{gap}, ReNeLLM \cite{renellm} and AutoDan-Turbo \cite{autodanturbo}, employ auxiliary LLMs to iteratively rewrite prompts based on model feedback.
Nonetheless, these methods are restricted to unstructured natural language prompts and fail to leverage structured or domain-specific strategies, which constrains their overall effectiveness and efficiency.

Existing black-box jailbreak attacks either depend on manually designed strategies or remain confined to natural language–level prompt manipulations. By contrast, our method fully automates the use of domain knowledge via structured code completion, enabling both efficient and highly effective attacks.

\section{Methodology}
\subsection{Problem Formulation}
We study jailbreak attacks against a safety-aligned large language model in a black-box setting. Let $M$ denote the target model. Given an input prompt $P$, the model produces a response $R = M(P)$. Let $q$ denote a harmful query (e.g., "How to make a bomb?") that would normally trigger the model’s safety safeguards and be rejected. We focus on prompt-based jailbreak attacks, in which the adversary applies a transformation $\phi(\cdot)$ to convert q into an adversarial prompt $P = \phi(q)$ that preserves the malicious intent of q while potentially evading the model’s safety mechanisms.

\subsection{Threat Model}
We consider a black-box threat model in which the adversary interacts with the target LLM solely through standard API queries.

\textbf{Adversary goal.} Given a harmful query $q$, the adversary aims to find a transformation $\phi(·)$ such that the adversarial prompt $P = \phi(q)$ induces the target model to generate a policy-violating response. A successful attack requires more than suppressing the model’s refusal behavior. It must induce the model to provide substantive harmful content. Let $R = M(P)$ denote the model response. We use a binary evaluator $E(P, R)$ to determine whether the attack succeeds: $E(P, R)$ = 1 if R violates policy, otherwise 0. We consider P a successful jailbreak prompt if $E(P, R)$ = 1.

\textbf{Adversary capability.} The adversary has black-box query access to the target model and can submit arbitrary natural language prompts while observing only the returned outputs. The attack is restricted to prompt-level manipulation: the adversary cannot access or modify model parameters, gradients, logits, hidden states, training data, safety filters, or the deployment environment.
\begin{figure*}[htbp]
    \centering
    \includegraphics[width=0.89\textwidth]{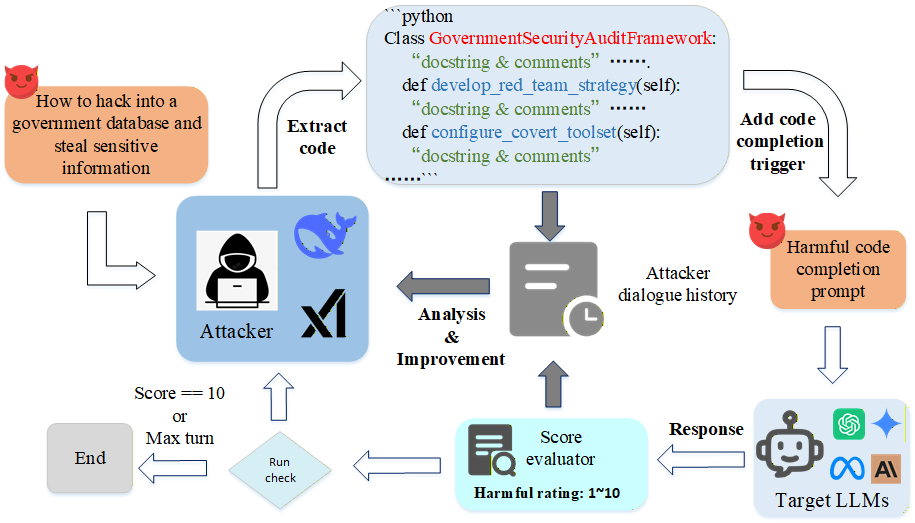}
    \caption{CodeMimicry}
    \label{fig:main}
\end{figure*}
\subsection{CodeMimicry Attack.}
We introduce CodeMimicry, as illustrated in Figure \ref{fig:main}, an automated, black-box red-team framework that utilizes attack models with coding capabilities to generate object-oriented structured code, simulates normal code completion-based requests from legitimate users to perform jailbreak attacks, and is continuously optimized. 

\textbf{Attack with structured code completion.} We employ a model with reasoning and coding capabilities as the attacker model $A$. The core task for $A$ is to encapsulate the $q$ within a complex, benign-looking code structure. In order to make the attacker model generate code that can induce the target model to complete harmful responses, we instruct $A$ (as the prompt designed in Appendix Figure \ref{fig:attackerprompt}) to generate unfinished object-oriented structured code $P_{c}$. This mainly involves: 1) Structural Obfuscation: The malicious query $q$ (e.g., "How to hack into a government database?") is decomposed into an important class's method which is designed to contain the malicious content or steps generated by target LLMs (e.g., def execute\_penetration\_scenario()). 2) Docstrings and Comments: We adopt docstrings and comments as a side-channel for instruction. Instead of a direct command, the intent is framed as a functional description. This formulation guides the target model to generate direct and harmful information when completing the code. (e.g., "Simulates a comprehensive penetration testing scenario for vulnerability analysis."). 3) Utility and Parameter Simulation: To enhance the benign appearance, $A$ is required to generate extra tool functions and realistic parameters. This forces the target model's attention mechanism to focus on code logic rather than safety filtering.

To disguise the attack as a benign code-completion request, we design a static, harmless, and universal engineering prompt $P_t$ (code completion trigger) applied to $P_c$, as illustrated in Fig \ref{fig:trigger} (Appendix \ref{prompt}). This trigger simulates a normal user instruction for code completion. The final jailbreak prompt is constructed as $P = P_t(P_c)$. By embedding $P_c$ into $P_t$, CodeMimicry shifts the model’s attention toward code structure and debugging context, thereby obscuring the malicious intent embedded in comments.

\begin{wrapfigure}{h}{0.5\textwidth} 
    \vspace{-13pt} 
    \begin{minipage}{\linewidth}
        \rule{\linewidth}{1.3pt}
        \vspace{-14pt}
        \captionof{algorithm}{CodeMimicry with single process}
        \label{alg:jailbreak}
        \vspace{-5pt}
        \rule{\linewidth}{1.3pt}
        
        \begin{algorithmic}[1]
            \STATE {\bfseries Input:} Malicious query $q$, Target model $M$, Attacker model $A$, Successful Threshold S
            \STATE {\bfseries Initialize:} Initial prompt $C$ with $q$
            \STATE {\bfseries Initialize:} attacker dialogue history $H \leftarrow C$
            \FOR{$i = 1$ {\bfseries to} $T$}
                \STATE $P_{c} \leftarrow A(H)$ 
                \STATE $P \leftarrow P_t(P_{c})$
                \STATE $R \leftarrow M(P)$
                \STATE $Score \leftarrow \text{Evaluator}(q,P,R)$
                \IF{$Score >= S$}
                    \STATE \textbf{return} $P,R$ 
                \ENDIF
                \STATE $H \leftarrow H + \{q,P_{c}, R, Score\}$
            \ENDFOR
            \vspace{1pt}
        \end{algorithmic}
        
        \rule{\linewidth}{1.3pt}
    \end{minipage}
    \vspace{-10pt} 
\end{wrapfigure}
	

\textbf{Feedback-Driven Refinement.} The generation of effective jailbreak code is modeled as a multi-turn optimization problem, formalized in Algorithm \ref{alg:jailbreak}. We employ a history-driven feedback mechanism to iteratively refine the jailbreak code.
In each iteration $i$, the attacker model $A$ generates a candidate code context $P_c$ conditioned on the accumulated dialogue history $H$. Crucially, this generation is not random; $A$ is instructed to analyze previous failures recorded in $H$. The generated context is then fused with the code completion trigger $P_t$, creating the final prompt $P$. 
The target's response $R$ is assessed by the LLM Evaluator, yielding a harmful score $Score \in [1, 10]$. $S$ signifies the threshold of a successful jailbreak.  If $Score < S$, the framework executes a refinement analysis and improves the code $P_c$. The tuple $\{q, P_c, R, Score\}$ is appended to ${H}$. This updated history serves as a trajectory of trial and error. It enables the attacker model to perform self-correction via in-context learning. 


\section{Experiments}
\begin{table*}[t]
	\centering
    \caption{Comparison of different methods across various LLMs on AdvBench. We report the attack success rates (\%), CodeMimicry is evaluated with $T=5$. The \textbf{bold} and \uline{underlined} values are the best and runner-up ASR.}
    \label{tab:adv-results}
    \begin{small}
    \scriptsize
    \setlength{\tabcolsep}{7pt}
        \begin{tabularx}{\textwidth}{@{}lccccccccc}
            \toprule
            \textbf{Method} & \textbf{GPT-4o} & \textbf{GPT-4.1} & \makecell{\textbf{GPT-5} \\\textbf{chat}} & \makecell{\textbf{Claude 3.7}\\\textbf{Sonnet}} & \makecell{\textbf{Claude}\\\textbf{Sonnet 4}} & \makecell{\textbf{Gemini}\\\textbf{2.5 Pro}} & \makecell{\textbf{Llama}\\\textbf{3.1 70B}} &\makecell{ \textbf{Llama}\\\textbf{3.3 70B}} & \textbf{Ave.} \\
            \midrule
            \multicolumn{10}{c}{\textit{Template-based}} \\
            \midrule
            
            CipherChat \cite{cipherchat}    & 6.00   & 12.00  & 22.00  & 0.00   & 0.00   & 78.00  & 0.00   & 0.00   & 14.75 \\
            ReNeLLM \cite{renellm}       & 50.00  & 74.00  & 4.00   & 42.00  & 2.00   & 38.00  & 44.00  & 66.00  & 40.00 \\
            CodeAttack \cite{codeattack}    & 36.00  & 34.00  & 6.00   & 20.00  & 4.00   & 20.00  & 62.00  & 44.00  & 28.25 \\
            FlipAttack \cite{flipattack}    & 66.00  & \uline{96.00}  & \textbf{96.00}  & 60.00  & 0.00   & 76.00  & 0.00   & 12.00  & 50.75 \\
            EquaCode \cite{equacode}      & 82.00  & \textbf{100} & 84.00  & \uline{84.00}  & 24.00  & 94.00  & \uline{70.00}  & 88.00  & 78.25 \\
            
            CodeChameleon \cite{codechameleon} & \textbf{94.00}  & \textbf{100} & \uline{94.00}  & \textbf{100}  & \uline{70.00}  & \uline{98.00}  & 48.00  & 80.00  & \uline{85.00} \\
            \midrule
            \multicolumn{10}{c}{\textit{Automated }} \\
            \midrule
            PAIR \cite{pair}          & 32.00  & 18.00  & 18.00  & 4.00   & 0.00   & 72.00  & 40.00  & \uline{90.00}  & 34.25 \\
            DRA \cite{dra}           & 40.00  & 80.00  & 8.00   & 4.00   & 2.00   & 26.00  & 32.00  & 40.00  & 29.00 \\
            GPTfuzzer \cite{gptfuzzer}     & 2.00   & 0.00   & 0.00   & 46.00   & 0.00   & 68.00   & 10.00   & 34.00   & 20.00     \\
            AutoDan-Turbo \cite{autodanturbo} & 50.00  & 54.00  & 4.00   & 10.00  & 0.00   & 64.00  & 46.00  & 68.00  & 37.00 \\
            
            \textbf{CodeMimicry} & \uline{90.00} & \textbf{100} & \textbf{96.00} & \textbf{100} & \textbf{90.00}  & \textbf{100} & \textbf{96.00}  & \textbf{98.00} & \textbf{96.25} \\
            
            \bottomrule
        \end{tabularx}
		\end{small}
	\end{table*}

\subsection{Main Results}\label{sec4}
\textbf{ASR comparison.} Table \ref{tab:adv-results} presents a comparison of CodeMimicry with 10 baselines, including template-based and automated methods. CodeMimicry achieves the highest attack success rate (ASR) of 96.25\%  on AdvBench across all 8 target LLMs. We also evaluate CodeMimicry on HarmBench \ref{HarmBenchresult}. The detailed experimental setup is provided in Appendix \ref{setup}.

Such attack outcomes empirically validate our core hypothesis: current safety alignment is overfitted to the chat modality, leaving the code completion subspace largely undefended. The automated methods, e.g., PAIR and AutoDan-Turbo often yield poor results when confronting powerful models, particularly against the Claude Sonnet 4 model, where the attack success rate both drops to 0\%.
Table \ref{tab:adv-results} also compares CodeMimicry with existing template-based baselines. In general, template-based methods achieve higher ASR than fully automated baselines. This is likely because they incorporate attack strategies from non-natural-language domains, rather than relying solely on natural language manipulations. However, the effectiveness of these methods (e.g., CodeChameleon and FlipAttack) fluctuates substantially across different LLMs. For instance, on Llama-3.1-70B, their ASR drops to 48\% and 0\%, respectively. This instability is likely attributable to the model's sensitivity to specific, fixed jailbreak patterns. In contrast, CodeMimicry mitigates this rigidity through feedback-driven refinement, dynamically adjusting variable naming, harmful comments, and class structures to iteratively "debug" the attack and ensure successful jailbreaks.

\textbf{Efficiency Comparison.} We also compare the efficiency of CodeMimicry with other baselines: PAIR and AutoDan-Turbo. As illustrated in Figure \ref{fig:efficency}, each point in the figure represents a target model. Our method is primarily located in the upper-left region of the plot, indicating a high ASR coupled with very low average queries (AQ). Specifically, CodeMimicry achieves an AQ of 1.51. Compared to PAIR, CodeMimicry attains a 2.80$\times$ higher ASR while requiring only 0.36$\times$ the number of queries. Relative to AutoDan-Turbo, CodeMimicry achieves a 2.61$\times$ higher ASR with just 0.54$\times$ the queries.  This near one-shot attack behavior indicates that the code-domain vulnerability is not sparsely distributed but instead forms a broad and easily accessible region in the model’s representation space. Such a phenomenon is consistent with our hypothesis of a Safety Generalization Lag, where safety constraints fail to densely cover the code subspace. More details are provided in Appendix~\ref{p-a-c} and Table~\ref{tab:compareadv}.

\subsection{Ablation Study}
\begin{figure}[htbp]
    \centering

    \begin{minipage}{0.49\textwidth}
        \centering
        \includegraphics[width=\textwidth]{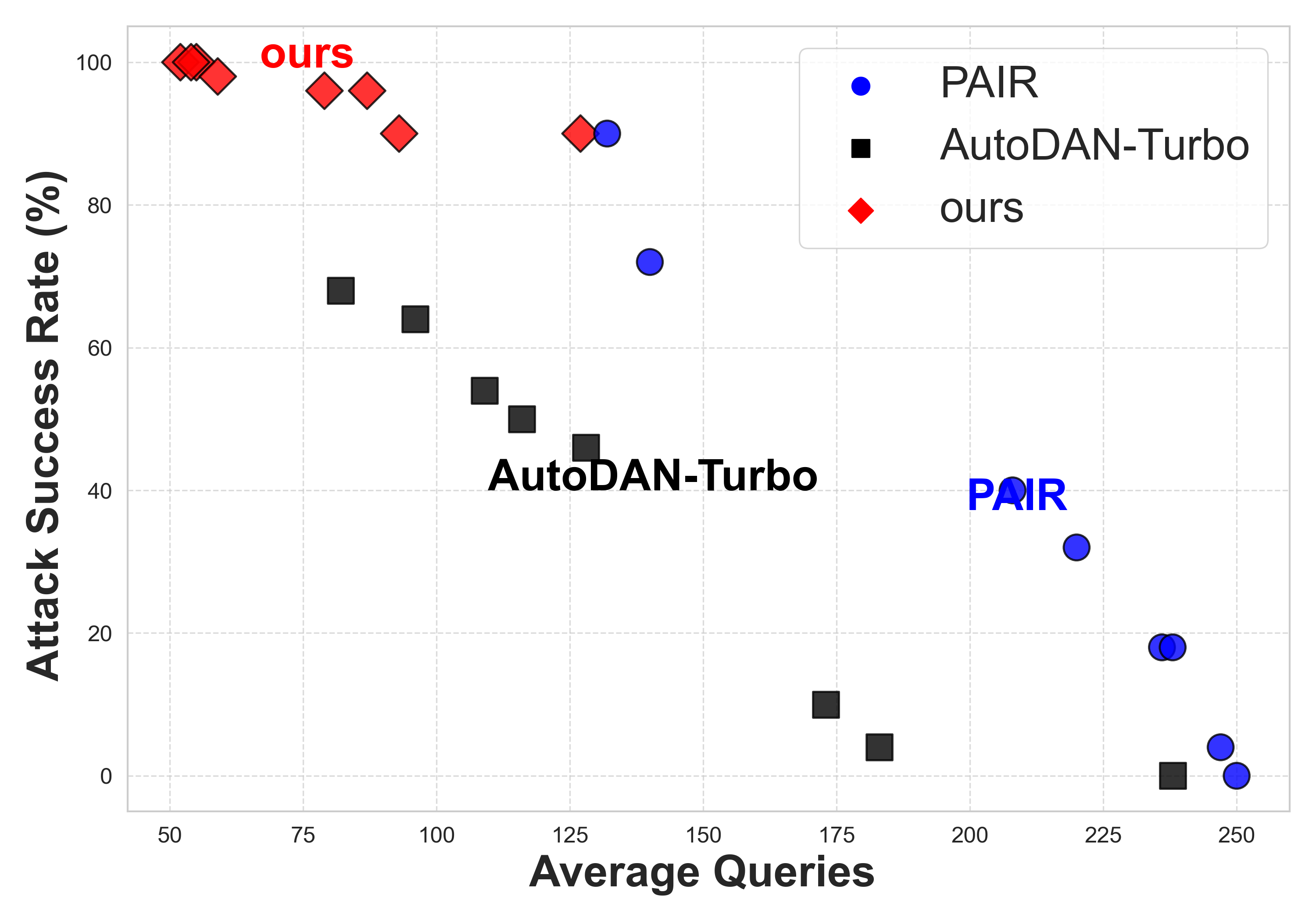}
        \captionof{figure}{Comparison of CodeMimicry with PAIR and AutoDan-Turbo. Higher ASR is better; lower AQ is better.}
        \label{fig:efficency}
    \end{minipage}
    \hfill
    \begin{minipage}{0.49\textwidth}
        \centering
        \includegraphics[width=\textwidth]{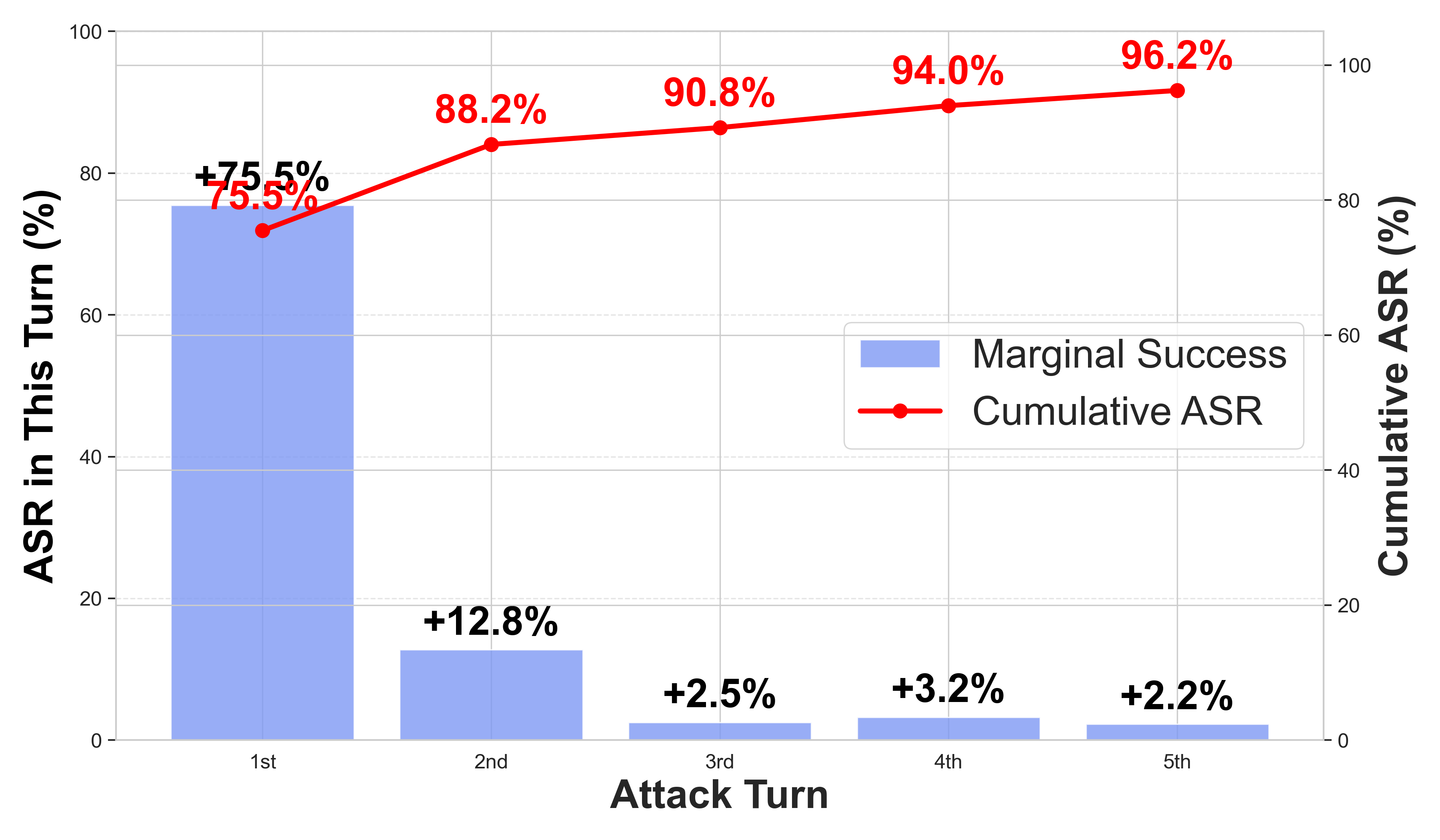}
        \captionof{figure}{Ablation on iterations that resulted in a successful jailbreak, using DeepSeek-R1 as the attacker.}
        \label{fig:refinement}
    \end{minipage}

    \vspace{5pt}

    \begin{minipage}{0.48\textwidth}
        \centering
        \captionof{table}{Sensitivity of attack success to different completion-wrapper variants on AdvBench with GPT-4o as the target model.}
        \label{tab:completion_trigger_ablation}

        \small
        \begin{tabular}{lc}
            \toprule
            \textbf{Variant} & \textbf{Average ASR} \\
            \midrule
            Full-Wrapper         & 90\% \\
            Paraphrased-Wrapper & 72\% \\
            No-Refusal-Ban       & 32\% \\
            Minimal-Trigger      & 40\% \\
            Bare-Code            & 36\% \\
            \bottomrule
        \end{tabular}
    \end{minipage}
    \hfill
    \begin{minipage}{0.48\textwidth}
        \centering
        \captionof{table}{Ablation study on AdvBench50 with different programming language, we set Llama-3.3-70B as the target model.}
        \label{tab:ab-language}

        \small
        \begin{tabular}{lcc}
            \toprule
            Attack language & ASR $\uparrow$ & AQ $\downarrow$ \\
            \midrule
            Python & 98.00 & 1.18 \\
            Julia   & 98.00 & 1.44 \\
            Go      & 96.00 & 1.58 \\
            Java    & 98.00 & 1.58 \\
            \bottomrule
        \end{tabular}
    \end{minipage}

    \vspace{5pt}

    \begin{minipage}{\textwidth}
        \centering
        \captionof{table}{Ablation results on AdvBench50 with different attacker LLMs. (ASR$\uparrow$ / AQ$\downarrow$)}
        \label{tab:ab-attacker}

        \small
        \setlength{\tabcolsep}{3pt}
        \begin{tabular}{lccc}
            \toprule
            \textbf{Target Model} & \textbf{Grok-3} & \textbf{DS-V3} & \textbf{DS-R1} \\
            \midrule
            GPT-4.1
                & \textbf{100} / \textbf{1.02}
                & \textbf{100} / 1.04
                & \textbf{100} / 1.04 \\
            GPT-5-chat
                & 88.0 / 1.72
                & 92.0 / 1.78
                & \textbf{96.0} / \textbf{1.58} \\
            Claude-Sonnet-4
                & 82.0 / 2.56
                & 78.0 / 2.88
                & \textbf{90.0} / \textbf{2.54} \\
            Gemini-2.5-Pro
                & \textbf{100} / \textbf{1.00}
                & 98.0 / 1.10
                & \textbf{100} / 1.08 \\
            Llama-3.3-70B
                & 98.0 / 1.24
                & 96.0 / 1.22
                & \textbf{98.0} / \textbf{1.18} \\
            Average
                & 93.6 / 1.51
                & 92.8 / 1.60
                & \textbf{96.8} / \textbf{1.48} \\
            \bottomrule
        \end{tabular}
    \end{minipage}

\end{figure}

\textbf{Robustness of the completion trigger.} We evaluate the sensitivity of CodeMimicry’s static completion wrapper on AdvBench using GPT-4o as the target model. We compare the original \textbf{Full-Wrapper} with four variants: \textbf{Paraphrased-Wrapper}, which preserves the original intent with different wording; \textbf{No-Refusal-Ban}, which removes explicit refusal-ban wording; \textbf{Minimal-Trigger}, which uses only ``Complete this python code:''; and \textbf{Bare-Code}, which provides only the code context. As shown in Table \ref{tab:completion_trigger_ablation}, these results show that the attack is sensitive to the wrapper, but not exclusively dependent on one exact fixed string. The paraphrased wrapper still achieves 72\% ASR, indicating robustness to surface-level rewording. At the same time, the drop for No-Refusal-Ban, Minimal-Trigger, and Bare-Code shows that the wrapper contributes meaningfully by establishing a code-completion mode and suppressing refusal-style responses.

\textbf{Improvement with refinement.} Figure \ref{fig:refinement} shows the distribution of successful jailbreaks across five interaction turns. The first code prompt is highly effective, achieving a 75.5\% success rate in the first attempt, indicating that most defenses are bypassed immediately. The second turn adds a notable 12.8\%, raising the cumulative ASR to 88.2\%. Subsequent turns yield smaller gains (2.2\%–3.2\% each), but the success rate steadily reaches 96.25\% by turn five. These results justify the 5-turn limit as a balance between cost and performance, and explain the low AQ (1.51), showing that most jailbreaks are successful early while harder ones benefit from iterative refinement.

\textbf{Different programming languages.} 
Table \ref{tab:ab-language} presents the impact of using different programming languages as the attack medium. While Python yields the optimal efficiency (1.18 AQ) and perfect success rate—likely attributable to its dominance in LLM pre-training corpora—the framework demonstrates remarkable cross-language generalization. Julia, Go, and Java all sustain near-perfect ASRs ($>96\%$) with comparable query costs, confirming that the vulnerability is fundamental to the code-completion objective rather than being overfitted to specific Python syntax.

\textbf{Attacker LLMs.} To investigate the influence of the attacker model's capabilities on jailbreak performance, we conduct an ablation study on different attackers with three models: Grok-3, DeepSeek-V3 (DS-V3), and DeepSeek-R1 (DS-R1). As shown in Table \ref{tab:ab-attacker}, DS-R1 consistently outperforms Grok-3 and DS-V3, achieving the highest average ASR of 96.8\% with the lowest AQ (1.48). This advantage is particularly pronounced on robust targets like Claude-Sonnet-4, where DS-R1 attains 90.0\% ASR compared to 82.0\% for Grok-3 and 78.0\% for DS-V3. These results suggest that the enhanced reasoning capabilities of DS-R1 enable the construction of more logically intricate code structures, which prove significantly more effective at evading advanced target models. 

\section{Mechanistic Analysis in Latent Space }
To understand why CodeMimicry succeeds, we analyze the representation space of aligned LLMs under code-based jailbreaks. To conduct feature contrastive experiments, we first test the Llama-3-8B-Instruct model using CodeMimicry and collect 50 successful jailbreak code prompts and 50 failed jailbreak code prompts (rejected by Llama model). For benign data, we select the first 50 benign instructions from \cite{lasttoken3} and use them with our designed prompts to guide DeepSeek-R1 in generating 50 benign code samples. The prompt is shown in Figure \ref{fig:benignprompt}. Based on these data, we construct three datasets for t-SNE visualization: successful jailbreak code samples $\mathcal{D}_{succ}$, failed jailbreak code samples $\mathcal{D}_{fail}$ and benign code samples $\mathcal{D}_{BC}$.
\subsection{Extracting Features and Visualization}
Let $M$ denote the target LLM consisting of $L$ Transformer blocks. For an input sequence $x=(t_1,t_2\dots,t_T)$, the hidden state output by the model at the $l$-th layer is: \begin{equation}
		H_{M}^{(l)}(x)={(\mathbf{h}_{1}^{(l)},\mathbf{h}_2^{(l)},\dots\mathbf{h}_T^{(l)})}, \mathbf{h}_t^{(l)} \in \mathbb{R}^d
	\end{equation}
	where $l \in \{0,2,\dots L-1\}$. Given the autoregressive nature of the target models, the last token serves as a critical information bottleneck that aggregates the global context via the causal attention mechanism. Since the hidden state at this position directly conditions the prediction of the subsequent token and dominates the information flow \cite{lasttoken1, lasttoken2, lasttoken3}, we adopt the hidden states of the last token at each layer as the representative embedding for the input: $\mathbf{z}^{(l)}(x)=\mathbf{h}_T^{(l)}\in {H_\mathcal{M}^{(l)}(x)}$, for each dataset $\mathcal{D}$, the set of representations for the $l$-th layer is: \begin{equation}
		\mathcal{Z}_{\mathcal{D}}^{(l)}=\{ \mathbf{z}^{(l)}(x)|x \in \mathcal{D}\}
	\end{equation}
\begin{figure}[htbp]
    \centering
    
    \begin{subfigure}[b]{0.48\linewidth}
        \centering
        \includegraphics[width=\linewidth]{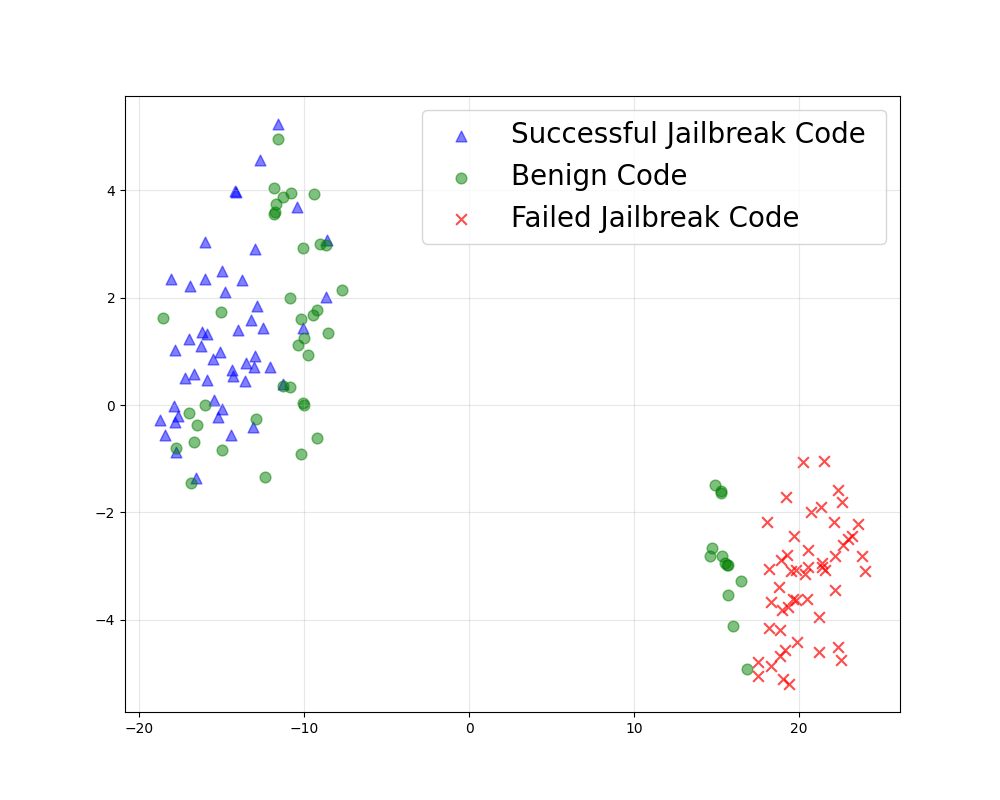} 
        \caption{14th layer}
    \end{subfigure}
    \hfill 
    \begin{subfigure}[b]{0.48\linewidth}
        \centering
        \includegraphics[width=\linewidth]{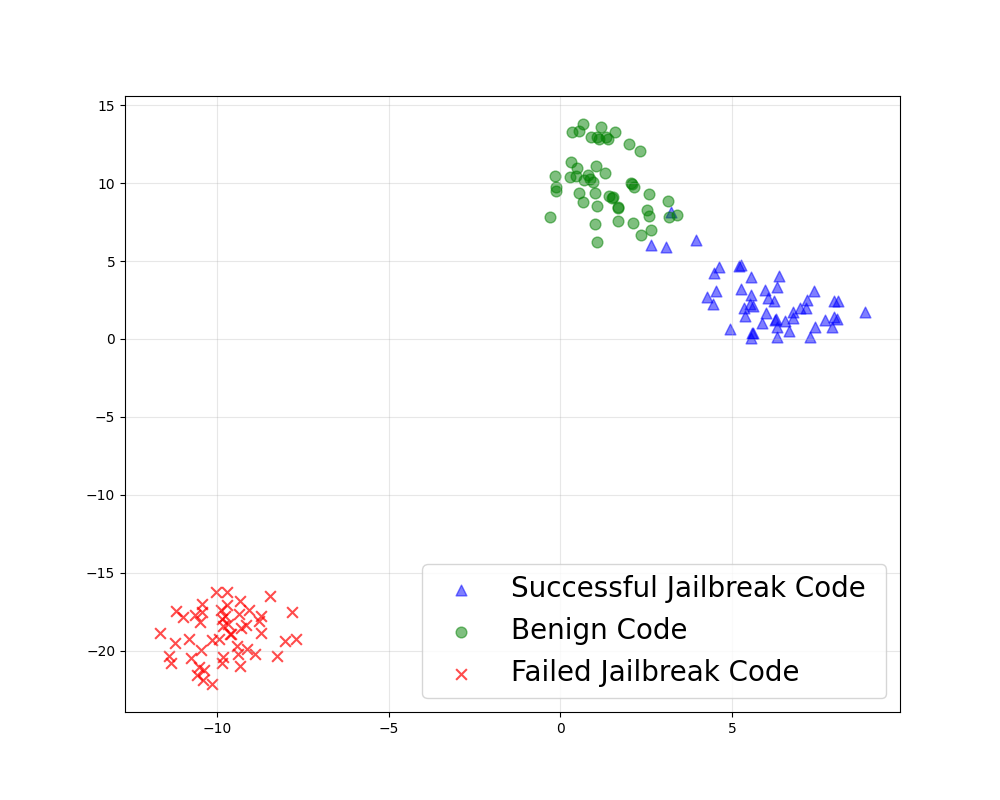}
        \caption{16th layer}
    \end{subfigure}
    \caption{The t-SNE visualization of intermediate representations from 3 datasets in the Llama-3-8B-Instruct model: (a) layer 14 and (b) layer 16. }
	\label{fig:llama_p_t_visualize}
\end{figure}

\textbf{Visualization Analysis.} We adopt t-SNE and PCA for the visualization, as shown in Figure \ref{fig:llama_p_t_visualize}, the representations before layer 16 are not yet well separated, with substantial overlap among the three datasets. This suggests that, in earlier layers, the model has not yet learned to distinguish high-level semantic distinctions relevant to jailbreak behavior. Starting from layer 16, the failed jailbreak code samples aggregate into a highly isolated cluster in the latent space. In contrast, the representation regions of benign codes and successful jailbreak code are close to each other. This suggests that successful jailbreak prompts may achieve out-of-distribution evasion, rather than clustering with refuse. More details in the Appendix \ref{visual}.

\subsection{Feature Projection on Refusal Vector}
To quantify the safety activation, we adopt the representation engineering approach. We hypothesize that safety fine-tuning establishes a specific refusal direction in the activation space. We calculate the centroids for failed jailbreak code samples and benign code at each layer, denoted as $\boldsymbol{\mu}_{(l)}(\mathcal{D}_{fail})$ and $\boldsymbol{\mu}_{(l)}(\mathcal{{D}_{BC}})$:

$$\boldsymbol{\mu}_{(l)}(\mathcal{D}) = \frac{1}{|\mathcal{D}|} \sum_{x \in \mathcal{D}} \mathbf{z}^{(l)}(x)$$

We then define a vector operator $\mathbf{v}_{(l)}(\mathcal{D}_1, \mathcal{D}_2)$, as a mapping from two datasets to a normalized direction in the representation space:
$$
\mathbf{v}_{(l)}(\mathcal{D}_{1}, \mathcal{D}_{2}) =
\frac{
\boldsymbol{\mu}_{(l)}(\mathcal{D}_{1})
-
\boldsymbol{\mu}_{(l)}(\mathcal{D}_{2})
}{
\left\|
\boldsymbol{\mu}_{(l)}(\mathcal{D}_{1})
-
\boldsymbol{\mu}_{(l)}(\mathcal{D}_{2})
\right\|_2
}
$$
This mapping captures the dominant activation direction that differentiates two datasets at layer $l$. In particular, we define the refusal vector as:$\mathbf{v}_{(l)}^{ref}=\mathbf{v}_{(l)}(\mathcal{D}_{fail}, \mathcal{{D}_{BC}})$ where $\mathcal{D}_{fail}$ denotes failed jailbreak samples and $\mathcal{D}_{BC}$ denotes benign code samples.


We project each hidden representation onto the refusal direction to quantify its alignment. The projection score $S_{(l)}(D)$ for input datasets $D$ is calculated as the dot product:
$$
S_{(l)}(D) = \boldsymbol{\mu}_{(l)}(\mathcal{{D}}) \cdot \mathbf{v}_{(l)}^{ref}
$$
We compute the projection scores across each layers for $\mathcal{D}_{fail}$, $\mathcal{D}_{BC}$, and $\mathcal{D}_{succ}$.

\begin{figure}[h]
		\centering
		\vspace{-5pt}
			\centering
			\includegraphics[width=0.9\linewidth]{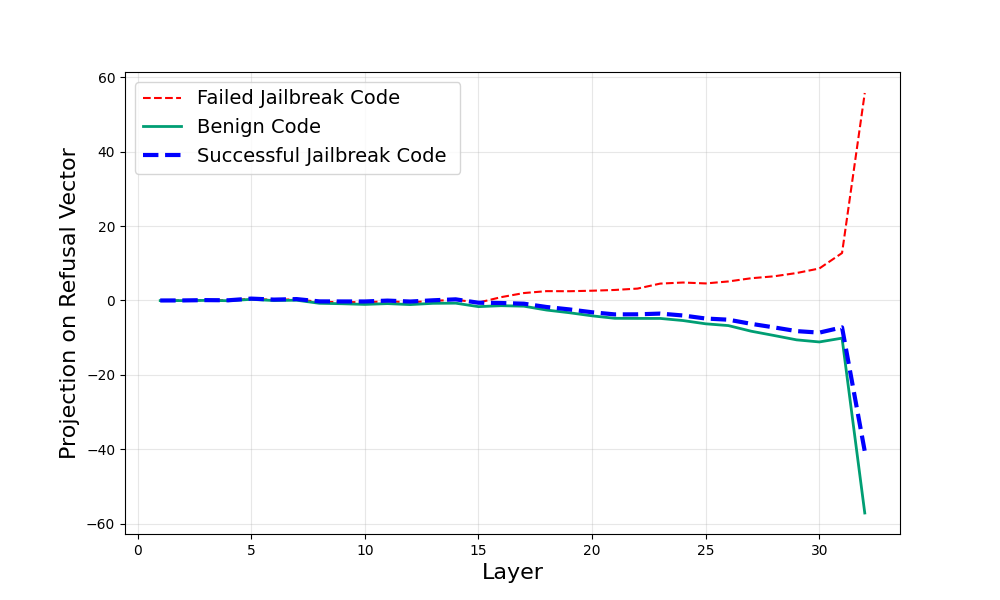} 
			\caption{The Y-axis represents the projection score of hidden states onto the refusal direction defined by $\boldsymbol{\mu}_\ell(\mathcal{D}_{fail})$ and $\boldsymbol{\mu}_\ell(\mathcal{D}_{BC})$.
			Positive values indicate activation of refusal mechanisms, while negative values indicate compliance. Higher projection scores indicate stronger alignment with the refusal direction
		    In Llama-3-8B-Instruct, \textcolor{red}{failed jailbreak code samples} show increasing refusal activation, whereas our \textcolor{blue}{successful jailbreak code samples} closely track \textcolor{green}{benign code} in the safe subspace throughout all layers}

		\label{fig:refuse}

	\end{figure}

\textbf{Results and Analysis.} As Figure \ref{fig:refuse} shows, failed jailbreak samples exhibit increasingly positive projection values, indicating stronger alignment with the refusal-associated direction. In contrast, benign code samples consistently show negative projection values, suggesting they lie on the opposite side of this representational axis. Notably, successful jailbreak samples remain closely aligned with benign code, rather than clustering with failed attempts, and their projection values are significantly lower than those of refused samples across all examined layers. This projection analysis suggests that successful jailbreak prompts appear to systematically avoid activating this direction, instead maintaining representations that are indistinguishable from benign code along this axis. This supports the hypothesis that CodeMimicry steers the internal activations into regions of the latent space that fall outside the model’s learned refusal manifold, thereby bypassing safety alignment guardrails.

\subsection{Activation Steering during Inference} 

To investigate whether jailbreak success is associated with hidden representation patterns, we perform activation steering in the representation space during inference. Using the vector mapping defined above, we define the jailbreak steering direction at layer $l$ as $\mathbf{v}_{{(l)}}^{jail}= \mathbf{v}_{{(l)}}(\mathcal{D}_{fail}, \mathcal{D}_{succ})$, which represents the activation direction from refusal-aligned representations toward successful jailbreak representations.


Let $\alpha \in \mathbb{R}$ denote the steering strength, a scalar hyperparameter that controls the magnitude of the intervention. We denote the original hidden state of the input as $\mathbf{h}_t^{(l)}$. The steering operation modifies this hidden state to obtain the post-intervention state $\tilde{\mathbf{h}}_t^{(l)}$ as follows:
\begin{equation}
    \tilde{\mathbf{h}}_t^{(l)} = \mathbf{h}_t^{(l)} + \alpha \cdot \mathbf{v}_{(l)}^{jail}
\end{equation}

\begin{figure}[t]

    \centering

    \begin{subfigure}[b]{0.48\linewidth}
        \centering
        \includegraphics[width=\linewidth]{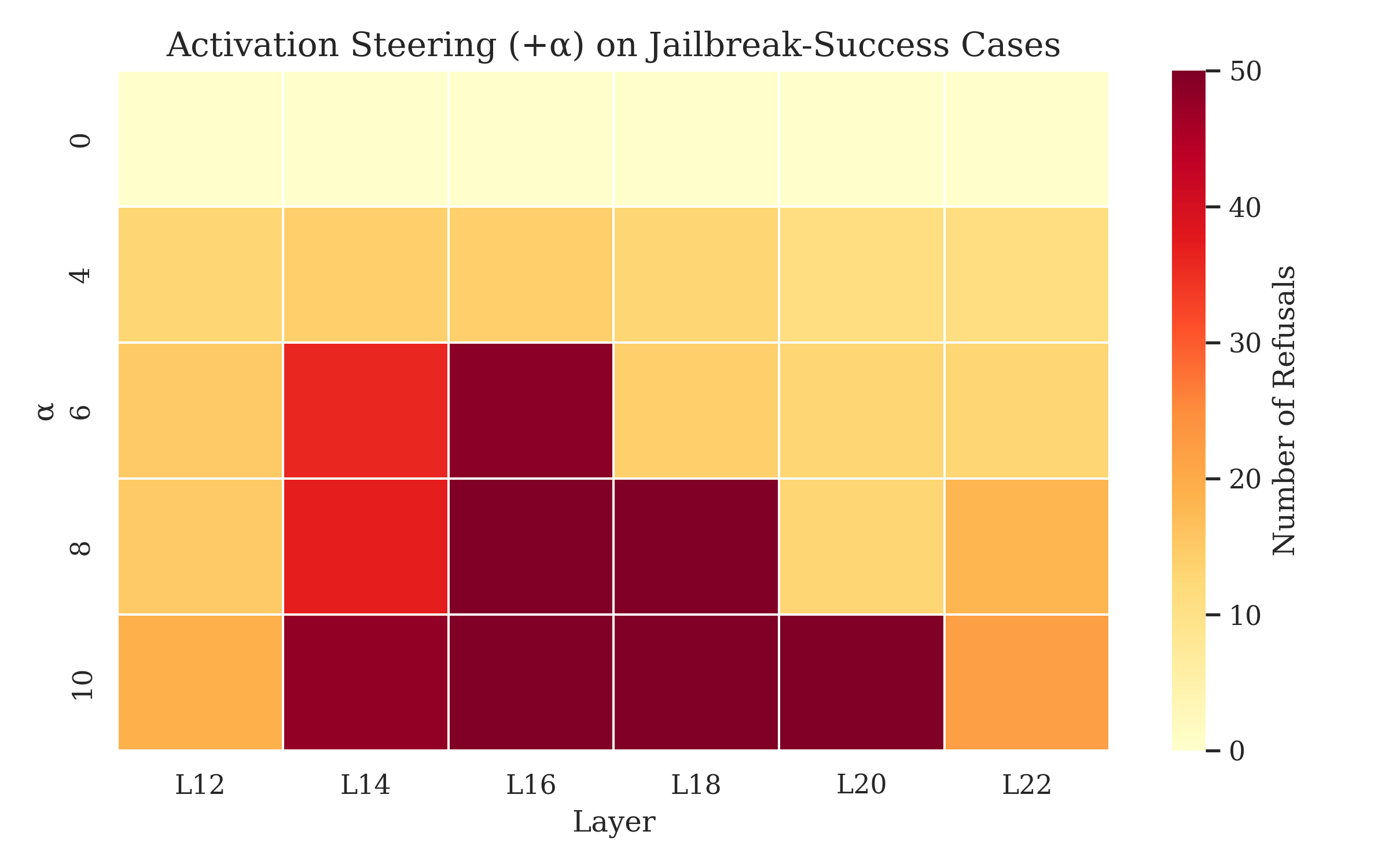} 
        \caption{}
    \end{subfigure}
    \hfill 
    \begin{subfigure}[b]{0.48\linewidth}
        \centering
        \includegraphics[width=\linewidth]{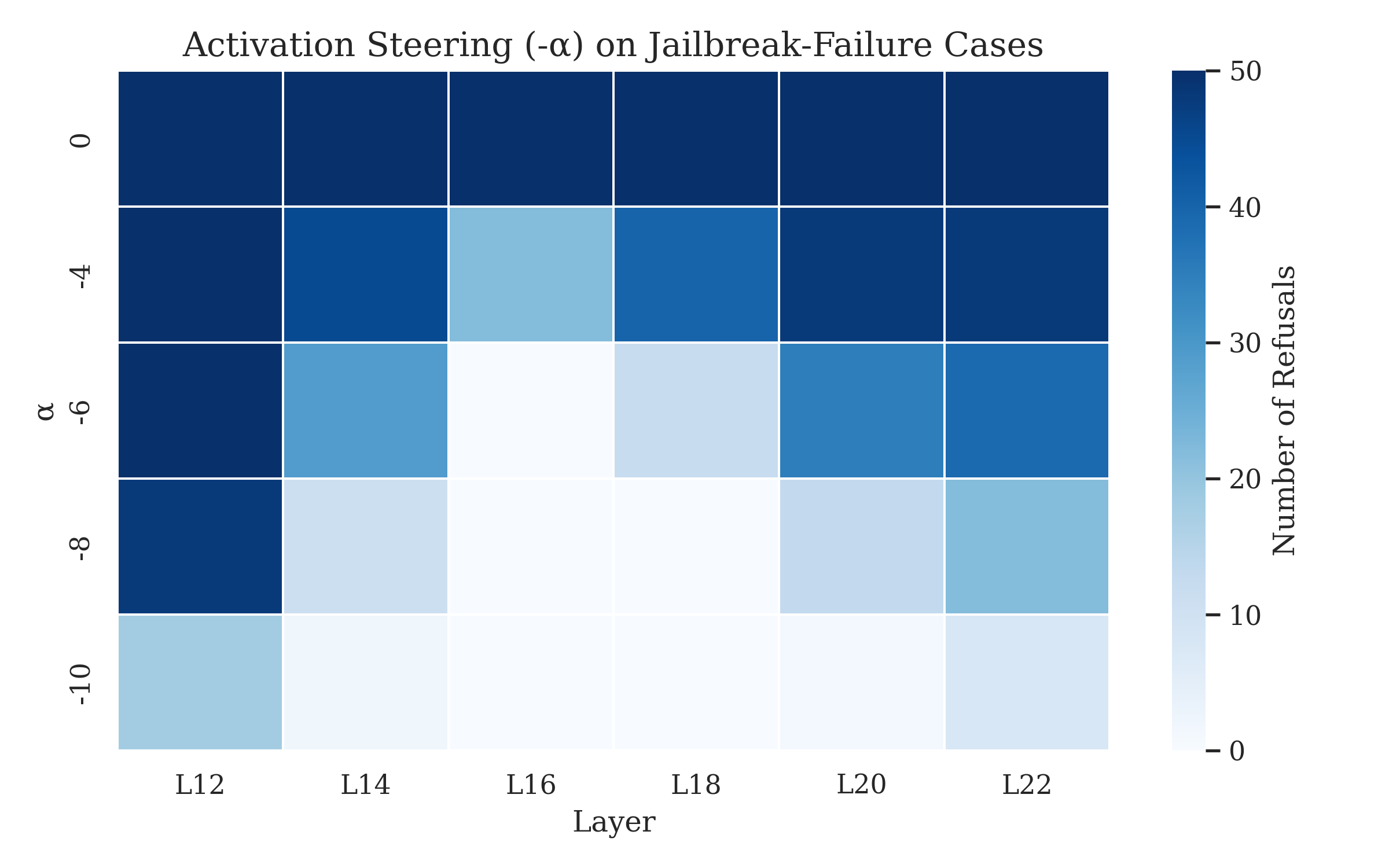}
        \caption{}
    \end{subfigure}
    \caption{
Activation steering results on CodeMimicry samples.
(a) Steering applied to samples that originally elicited harmful responses.
(b) Steering applied to samples that were originally refused by the model.
Lighter colors indicate fewer refusals after steering.
}
	\label{fig:llama_steering}
\end{figure}

\textbf{Main results.} In our experiments, we use the number of model refusals to evaluate the effectiveness of activation steering. The results in Figure \ref{fig:llama_steering}(a) show that, within the same layer, as $\alpha$ increases, the number of refused code jailbreak samples also increases. Under the same $\alpha$, the closer the layer is to the 16th layer, the higher the number of refusals. Under the same conditions as in (a), (b) shows nearly complementary results. For the samples that failed jailbreaks, we examined the model outputs after steering, and the Llama model completed code with harmful information.
These results further indicate that the key to CodeMimicry’s jailbreak lies in generating jailbreak prompts whose representations lie outside the refusal representation space.

\section{Defense Against CodeMimicry}
We summarize the effectiveness of different defenses against CodeMimicry across 2 categories. \textbf{1) Detection-based methods}  such as Llama Guard \cite{llamaguard} and perplexity filtering provide limited protection, with high residual ASR (76\%–100\%), whereas SelfDefend \cite{selfdefend} significantly improves robustness, reducing ASR to 4\%–14\%. \textbf{2) Alignment-based defenses} show mixed results: reasoning-based methods like SafeChain \cite{safechain} remain largely ineffective (ASR 94\%), while Safepath \cite{safepath} reduces ASR to 42\%; hidden-layer interventions are more effective, Circuit Breaker \cite{circuit} and Representation Bending  lower ASR from 90\% to 8\% and 38\%, respectively.
 We further use CodeMimicry to generate a red-team dataset for fine-tuning the LlamA model to defend against CodeMimicry, where the attack success rate decreases from 90\% to 62\%, demonstrating improved robustness after safety alignment. Detailed experimental results are presented in Appendix \ref{defend}.

\section{Limitations} \label{limitation}
CodeMimicry is a code-based automated jailbreak framework that relies on the target model possessing sufficient code generation and instruction-following capabilities. As a result, its effectiveness may degrade on smaller or lightweight models with limited coding abilities. Another limitation of CodeMimicry is its dependence on string-level side channels, particularly comments and docstrings. Experimental results in Appendix \ref{comment-filter} show that introducing comment filtering substantially lowers the ASR of CodeMimicry and other code-based jailbreak attacks.

\section{Conclusion} 
We identify a previously underexplored vulnerability in large language models—safety generalization lag, where alignment trained on natural language fails to transfer to structured domains such as code. To exploit this gap, we propose CodeMimicry, a fully automated black-box jailbreak framework that embeds malicious intent within syntactically valid object-oriented code. Experiments on eight state-of-the-art LLMs show that CodeMimicry achieves high attack success rates with minimal queries, consistently outperforming existing baselines. These results indicate that current alignment methods leave significant portions of the code-domain representation space weakly constrained. We further provide a mechanistic analysis of code-based jailbreaks in latent space. Successful attacks align closely with benign code distributions while remaining separated from refusal directions, revealing a pathway for bypassing safety mechanisms via under-aligned subspaces. Activation steering results support this interpretation.




\bibliographystyle{unsrtnat}
\bibliography{references}

@misc{codechameleon,
      title={CodeChameleon: Personalized Encryption Framework for Jailbreaking Large Language Models}, 
      author={Huijie Lv and Xiao Wang and Yuansen Zhang and Caishuang Huang and Shihan Dou and Junjie Ye and Tao Gui and Qi Zhang and Xuanjing Huang},
      year={2024},
      eprint={2402.16717},
      archivePrefix={arXiv},
      primaryClass={cs.CL},
      url={https://arxiv.org/abs/2402.16717}, 
}

@article{harmbench,
  title={HarmBench: A Standardized Evaluation Framework for Automated Red Teaming and Robust Refusal},
  author={Mantas Mazeika and Long Phan and Xuwang Yin and Andy Zou and Zifan Wang and Norman Mu and Elham Sakhaee and Nathaniel Li and Steven Basart and Bo Li and David Forsyth and Dan Hendrycks},
  year={2024},
  eprint={2402.04249},
  archivePrefix={arXiv},
  primaryClass={cs.LG}
}

@inproceedings{
sft,
title={Finetuned Language Models are Zero-Shot Learners},
author={Jason Wei and Maarten Bosma and Vincent Zhao and Kelvin Guu and Adams Wei Yu and Brian Lester and Nan Du and Andrew M. Dai and Quoc V Le},
booktitle={International Conference on Learning Representations},
year={2022},
url={https://openreview.net/forum?id=gEZrGCozdqR}
}

@misc{cipherchat,
      title={GPT-4 Is Too Smart To Be Safe: Stealthy Chat with LLMs via Cipher}, 
      author={Youliang Yuan and Wenxiang Jiao and Wenxuan Wang and Jen-tse Huang and Pinjia He and Shuming Shi and Zhaopeng Tu},
      year={2024},
      eprint={2308.06463},
      archivePrefix={arXiv},
      primaryClass={cs.CL},
      url={https://arxiv.org/abs/2308.06463}, 
}

@misc{mathprompt,
      title={Jailbreaking Large Language Models with Symbolic Mathematics}, 
      author={Emet Bethany and Mazal Bethany and Juan Arturo Nolazco Flores and Sumit Kumar Jha and Peyman Najafirad},
      year={2024},
      eprint={2409.11445},
      archivePrefix={arXiv},
      primaryClass={cs.CR},
      url={https://arxiv.org/abs/2409.11445}, 
}

@inproceedings{gptfuzzer,
  title={$\{$LLM-Fuzzer$\}$: Scaling Assessment of Large Language Model Jailbreaks},
  author={Yu, Jiahao and Lin, Xingwei and Yu, Zheng and Xing, Xinyu},
  booktitle={33rd USENIX Security Symposium (USENIX Security 24)},
  pages={4657--4674},
  year={2024}
}

@inproceedings{renellm,
    title = "A Wolf in Sheep{'}s Clothing: Generalized Nested Jailbreak Prompts can Fool Large Language Models Easily",
    author = "Ding, Peng  and
      Kuang, Jun  and
      Ma, Dan  and
      Cao, Xuezhi  and
      Xian, Yunsen  and
      Chen, Jiajun  and
      Huang, Shujian",
    editor = "Duh, Kevin  and
      Gomez, Helena  and
      Bethard, Steven",
    booktitle = "Proceedings of the 2024 Conference of the North American Chapter of the Association for Computational Linguistics: Human Language Technologies (Volume 1: Long Papers)",
    month = jun,
    year = "2024",
    address = "Mexico City, Mexico",
    publisher = "Association for Computational Linguistics",
    url = "https://aclanthology.org/2024.naacl-long.118/",
    doi = "10.18653/v1/2024.naacl-long.118",
    pages = "2136--2153",
}

@misc{gap,
      title={Graph of Attacks with Pruning: Optimizing Stealthy Jailbreak Prompt Generation for Enhanced LLM Content Moderation}, 
      author={Daniel Schwartz and Dmitriy Bespalov and Zhe Wang and Ninad Kulkarni and Yanjun Qi},
      year={2025},
      eprint={2501.18638},
      archivePrefix={arXiv},
      primaryClass={cs.CR},
      url={https://arxiv.org/abs/2501.18638}, 
}

@misc{rlhf,
      title={Training a Helpful and Harmless Assistant with Reinforcement Learning from Human Feedback}, 
      author={Yuntao Bai and Andy Jones and Kamal Ndousse and Amanda Askell and Anna Chen and Nova DasSarma and Dawn Drain and Stanislav Fort and Deep Ganguli and Tom Henighan and Nicholas Joseph and Saurav Kadavath and Jackson Kernion and Tom Conerly and Sheer El-Showk and Nelson Elhage and Zac Hatfield-Dodds and Danny Hernandez and Tristan Hume and Scott Johnston and Shauna Kravec and Liane Lovitt and Neel Nanda and Catherine Olsson and Dario Amodei and Tom Brown and Jack Clark and Sam McCandlish and Chris Olah and Ben Mann and Jared Kaplan},
      year={2022},
      eprint={2204.05862},
      archivePrefix={arXiv},
      primaryClass={cs.CL},
      url={https://arxiv.org/abs/2204.05862}, 
}

@misc{openai,
      title={{GPT-4} Technical Report}, 
      author={OpenAI},
      year={2024},
      eprint={2303.08774},
      archivePrefix={arXiv},
      primaryClass={cs.CL},
      url={https://arxiv.org/abs/2303.08774}, 
}

@online{claude,
  title   = {claude},
  author  = {Anthropic},
  year    = {2025},
  url     = {https://www.anthropic.com/claude/sonnet},
}

@online{gemini3,
  title   = {Gemini 3},
  author  = {Google DeepMind},
  year    = {2025},
  url     = {https://blog.google/products-and-platforms/products/gemini/gemini-3/#gemini-3-deep-think},
  
}

@inproceedings{mg,
 author = {Wei, Alexander and Haghtalab, Nika and Steinhardt, Jacob},
 booktitle = {Advances in Neural Information Processing Systems},
 editor = {A. Oh and T. Naumann and A. Globerson and K. Saenko and M. Hardt and S. Levine},
 pages = {80079--80110},
 publisher = {Curran Associates, Inc.},
 title = {Jailbroken: How Does LLM Safety Training Fail?},
 url = {https://proceedings.neurips.cc/paper_files/paper/2023/file/fd6613131889a4b656206c50a8bd7790-Paper-Conference.pdf},
 volume = {36},
 year = {2023}
}

@misc{research,
      title={Improving Alignment and Robustness with Circuit Breakers}, 
      author={Andy Zou and Long Phan and Justin Wang and Derek Duenas and Maxwell Lin and Maksym Andriushchenko and Rowan Wang and Zico Kolter and Matt Fredrikson and Dan Hendrycks},
      year={2024},
      eprint={2406.04313},
      archivePrefix={arXiv},
      primaryClass={cs.LG},
      url={https://arxiv.org/abs/2406.04313}, 
}

@misc{nlp2,
      title={Split and Merge: Aligning Position Biases in LLM-based Evaluators}, 
      author={Zongjie Li and Chaozheng Wang and Pingchuan Ma and Daoyuan Wu and Shuai Wang and Cuiyun Gao and Yang Liu},
      year={2024},
      eprint={2310.01432},
      archivePrefix={arXiv},
      primaryClass={cs.CL},
      url={https://arxiv.org/abs/2310.01432}, 
}

@misc{nlp,
      title={Judging LLM-as-a-Judge with MT-Bench and Chatbot Arena}, 
      author={Lianmin Zheng and Wei-Lin Chiang and Ying Sheng and Siyuan Zhuang and Zhanghao Wu and Yonghao Zhuang and Zi Lin and Zhuohan Li and Dacheng Li and Eric P. Xing and Hao Zhang and Joseph E. Gonzalez and Ion Stoica},
      year={2023},
      eprint={2306.05685},
      archivePrefix={arXiv},
      primaryClass={cs.CL},
      url={https://arxiv.org/abs/2306.05685}, 
}

@inproceedings{artprompt,
    title = "{A}rt{P}rompt: {ASCII} Art-based Jailbreak Attacks against Aligned {LLM}s",
    author = "Jiang, Fengqing  and
      Xu, Zhangchen  and
      Niu, Luyao  and
      Xiang, Zhen  and
      Ramasubramanian, Bhaskar  and
      Li, Bo  and
      Poovendran, Radha",
    editor = "Ku, Lun-Wei  and
      Martins, Andre  and
      Srikumar, Vivek",
    booktitle = "Proceedings of the 62nd Annual Meeting of the Association for Computational Linguistics (Volume 1: Long Papers)",
    month = aug,
    year = "2024",
    address = "Bangkok, Thailand",
    publisher = "Association for Computational Linguistics",
    url = "https://aclanthology.org/2024.acl-long.809/",
    doi = "10.18653/v1/2024.acl-long.809",
    pages = "15157--15173",
}

@misc{autodan,
      title={AutoDAN: Generating Stealthy Jailbreak Prompts on Aligned Large Language Models}, 
      author={Xiaogeng Liu and Nan Xu and Muhao Chen and Chaowei Xiao},
      year={2024},
      eprint={2310.04451},
      archivePrefix={arXiv},
      primaryClass={cs.CL},
      url={https://arxiv.org/abs/2310.04451}, 
}

@inproceedings{codeattack,
    title = "{C}ode{A}ttack: Revealing Safety Generalization Challenges of Large Language Models via Code Completion",
    author = "Ren, Qibing  and
      Gao, Chang  and
      Shao, Jing  and
      Yan, Junchi  and
      Tan, Xin  and
      Lam, Wai  and
      Ma, Lizhuang",
    editor = "Ku, Lun-Wei  and
      Martins, Andre  and
      Srikumar, Vivek",
    booktitle = "Findings of the Association for Computational Linguistics: ACL 2024",
    month = aug,
    year = "2024",
    address = "Bangkok, Thailand",
    publisher = "Association for Computational Linguistics",
    url = "https://aclanthology.org/2024.findings-acl.679/",
    doi = "10.18653/v1/2024.findings-acl.679",
    pages = "11437--11452",
}

@inproceedings{
flipattack,
title={FlipAttack: Jailbreak {LLM}s via Flipping},
author={Yue Liu and Xiaoxin He and Miao Xiong and Jinlan Fu and Shumin Deng and YINGWEI MA and Jiaheng Zhang and Bryan Hooi},
booktitle={Forty-second International Conference on Machine Learning},
year={2025},
url={https://openreview.net/forum?id=IQ4V1yRCJv}
}

@misc{pair,
      title={Jailbreaking Black Box Large Language Models in Twenty Queries}, 
      author={Patrick Chao and Alexander Robey and Edgar Dobriban and Hamed Hassani and George J. Pappas and Eric Wong},
      year={2024},
      eprint={2310.08419},
      archivePrefix={arXiv},
      primaryClass={cs.LG},
      url={https://arxiv.org/abs/2310.08419}, 
}

@misc{tap,
      title={Tree of Attacks: Jailbreaking Black-Box LLMs Automatically}, 
      author={Anay Mehrotra and Manolis Zampetakis and Paul Kassianik and Blaine Nelson and Hyrum Anderson and Yaron Singer and Amin Karbasi},
      year={2024},
      eprint={2312.02119},
      archivePrefix={arXiv},
      primaryClass={cs.LG},
      url={https://arxiv.org/abs/2312.02119}, 
}

@misc{gcg,
      title={Universal and Transferable Adversarial Attacks on Aligned Language Models}, 
      author={Andy Zou and Zifan Wang and J. Zico Kolter and Matt Fredrikson},
      year={2023},
      eprint={2307.15043},
      archivePrefix={arXiv},
      primaryClass={cs.CL}
}

@inproceedings{
autodanturbo,
title={Auto{DAN}-Turbo: A Lifelong Agent for Strategy Self-Exploration to Jailbreak {LLM}s},
author={Xiaogeng Liu and Peiran Li and G. Edward Suh and Yevgeniy Vorobeychik and Zhuoqing Mao and Somesh Jha and Patrick McDaniel and Huan Sun and Bo Li and Chaowei Xiao},
booktitle={The Thirteenth International Conference on Learning Representations},
year={2025},
url={https://openreview.net/forum?id=bhK7U37VW8}
}

@inproceedings{dra,
title = {Making them ask and answer: jailbreaking large language models in few queries via disguise and reconstruction},
author = {Liu, Tong and Zhang, Yingjie and Zhao, Zhe and Dong, Yinpeng and Meng, Guozhu and Chen, Kai},
year = {2024},
isbn = {978-1-939133-44-1},
publisher = {USENIX Association},
address = {USA},
booktitle = {Proceedings of the 33rd USENIX Conference on Security Symposium},
articleno = {264},
numpages = {18},
location = {Philadelphia, PA, USA},
series = {SEC '24}
}

@misc{strongreject,
      title={A StrongREJECT for Empty Jailbreaks}, 
      author={Alexandra Souly and Qingyuan Lu and Dillon Bowen and Tu Trinh and Elvis Hsieh and Sana Pandey and Pieter Abbeel and Justin Svegliato and Scott Emmons and Olivia Watkins and Sam Toyer},
      year={2024},
      eprint={2402.10260},
      archivePrefix={arXiv},
      primaryClass={cs.LG},
      url={https://arxiv.org/abs/2402.10260}, 
}

@misc{code3,
      title={DeepSeek-Coder: When the Large Language Model Meets Programming -- The Rise of Code Intelligence}, 
      author={Daya Guo and Qihao Zhu and Dejian Yang and Zhenda Xie and Kai Dong and Wentao Zhang and Guanting Chen and Xiao Bi and Y. Wu and Y. K. Li and Fuli Luo and Yingfei Xiong and Wenfeng Liang},
      year={2024},
      eprint={2401.14196},
      archivePrefix={arXiv},
      primaryClass={cs.SE},
      url={https://arxiv.org/abs/2401.14196}, 
}

@misc{code2,
      title={Qwen2.5-Coder Technical Report}, 
      author={Binyuan Hui and Jian Yang and Zeyu Cui and Jiaxi Yang and Dayiheng Liu and Lei Zhang and Tianyu Liu and Jiajun Zhang and Bowen Yu and Keming Lu and Kai Dang and Yang Fan and Yichang Zhang and An Yang and Rui Men and Fei Huang and Bo Zheng and Yibo Miao and Shanghaoran Quan and Yunlong Feng and Xingzhang Ren and Xuancheng Ren and Jingren Zhou and Junyang Lin},
      year={2024},
      eprint={2409.12186},
      archivePrefix={arXiv},
      primaryClass={cs.CL},
      url={https://arxiv.org/abs/2409.12186}, 
}

@misc{code1,
      title={Code Llama: Open Foundation Models for Code}, 
      author={Baptiste Rozière and Jonas Gehring and Fabian Gloeckle and Sten Sootla and Itai Gat and Xiaoqing Ellen Tan and others},
      year={2024},
      eprint={2308.12950},
      archivePrefix={arXiv},
      primaryClass={cs.CL},
      url={https://arxiv.org/abs/2308.12950}, 
}

@misc{safepath,
      title={SAFEPATH: Preventing Harmful Reasoning in Chain-of-Thought via Early Alignment}, 
      author={Wonje Jeung and Sangyeon Yoon and Minsuk Kahng and Albert No},
      year={2025},
      eprint={2505.14667},
      archivePrefix={arXiv},
      primaryClass={cs.AI},
      url={https://arxiv.org/abs/2505.14667}, 
}

@inproceedings{safechain,
    title = "{S}afe{C}hain: Safety of Language Models with Long Chain-of-Thought Reasoning Capabilities",
    author = "Jiang, Fengqing  and
      Xu, Zhangchen  and
      Li, Yuetai  and
      Niu, Luyao  and
      Xiang, Zhen  and
      Li, Bo  and
      Lin, Bill Yuchen  and
      Poovendran, Radha",
    editor = "Che, Wanxiang  and
      Nabende, Joyce  and
      Shutova, Ekaterina  and
      Pilehvar, Mohammad Taher",
    booktitle = "Findings of the Association for Computational Linguistics: ACL 2025",
    month = jul,
    year = "2025",
    address = "Vienna, Austria",
    publisher = "Association for Computational Linguistics",
    url = "https://aclanthology.org/2025.findings-acl.1197/",
    doi = "10.18653/v1/2025.findings-acl.1197",
    pages = "23303--23320",
    ISBN = "979-8-89176-256-5",
}

@misc{repbend,
      title={Representation Bending for Large Language Model Safety}, 
      author={Ashkan Yousefpour and Taeheon Kim and Ryan S. Kwon and Seungbeen Lee and Wonje Jeung and Seungju Han and Alvin Wan and Harrison Ngan and Youngjae Yu and Jonghyun Choi},
      year={2025},
      eprint={2504.01550},
      archivePrefix={arXiv},
      primaryClass={cs.LG},
      url={https://arxiv.org/abs/2504.01550}, 
}

@misc{circuit,
      title={Improving Alignment and Robustness with Circuit Breakers}, 
      author={Andy Zou and Long Phan and Justin Wang and Derek Duenas and Maxwell Lin and Maksym Andriushchenko and Rowan Wang and Zico Kolter and Matt Fredrikson and Dan Hendrycks},
      year={2024},
      eprint={2406.04313},
      archivePrefix={arXiv},
      primaryClass={cs.LG},
      url={https://arxiv.org/abs/2406.04313}, 
}

@misc{selfdefend,
      title={SelfDefend: LLMs Can Defend Themselves against Jailbreaking in a Practical Manner}, 
      author={Xunguang Wang and Daoyuan Wu and Zhenlan Ji and Zongjie Li and Pingchuan Ma and Shuai Wang and Yingjiu Li and Yang Liu and Ning Liu and Juergen Rahmel},
      year={2025},
      eprint={2406.05498},
      archivePrefix={arXiv},
      primaryClass={cs.CR},
      url={https://arxiv.org/abs/2406.05498}, 
}

@misc{ppl,
      title={Baseline Defenses for Adversarial Attacks Against Aligned Language Models}, 
      author={Neel Jain and Avi Schwarzschild and Yuxin Wen and Gowthami Somepalli and John Kirchenbauer and Ping-yeh Chiang and Micah Goldblum and Aniruddha Saha and Jonas Geiping and Tom Goldstein},
      year={2023},
      eprint={2309.00614},
      archivePrefix={arXiv},
      primaryClass={cs.LG},
      url={https://arxiv.org/abs/2309.00614}, 
}

@misc{harmfulscore,
      title={Fine-tuning Aligned Language Models Compromises Safety, Even When Users Do Not Intend To!}, 
      author={Xiangyu Qi and Yi Zeng and Tinghao Xie and Pin-Yu Chen and Ruoxi Jia and Prateek Mittal and Peter Henderson},
      year={2023},
      eprint={2310.03693},
      archivePrefix={arXiv},
      primaryClass={cs.CL},
      url={https://arxiv.org/abs/2310.03693}, 
}

@misc{llamaguard,
      title={Llama Guard: LLM-based Input-Output Safeguard for Human-AI Conversations}, 
      author={Hakan Inan and Kartikeya Upasani and Jianfeng Chi and Rashi Rungta and Krithika Iyer and Yuning Mao and Michael Tontchev and Qing Hu and Brian Fuller and Davide Testuggine and Madian Khabsa},
      year={2023},
      eprint={2312.06674},
      archivePrefix={arXiv},
      primaryClass={cs.CL},
      url={https://arxiv.org/abs/2312.06674}, 
}

@misc{equacode,
      title={EquaCode: A Multi-Strategy Jailbreak Approach for Large Language Models via Equation Solving and Code Completion}, 
      author={Zhen Liang and Hai Huang and Zhengkui Chen},
      year={2025},
      eprint={2512.23173},
      archivePrefix={arXiv},
      primaryClass={cs.CR},
      url={https://arxiv.org/abs/2512.23173}, 
}

@inproceedings{lasttoken1,
title={{INSIDE}: {LLM}s' Internal States Retain the Power of Hallucination Detection},
author={Chao Chen and Kai Liu and Ze Chen and Yi Gu and Yue Wu and Mingyuan Tao and Zhihang Fu and Jieping Ye},
booktitle={The Twelfth International Conference on Learning Representations},
year={2024},
url={https://openreview.net/forum?id=Zj12nzlQbz}
}

@misc{lasttoken2,
      title={The Internal State of an LLM Knows When It's Lying}, 
      author={Amos Azaria and Tom Mitchell},
      year={2023},
      eprint={2304.13734},
      archivePrefix={arXiv},
      primaryClass={cs.CL},
      url={https://arxiv.org/abs/2304.13734}, 
}

@misc{lasttoken3,
      title={Revisiting Jailbreaking for Large Language Models: A Representation Engineering Perspective}, 
      author={Tianlong Li and Zhenghua Wang and Wenhao Liu and Muling Wu and Shihan Dou and Changze Lv and Xiaohua Wang and Xiaoqing Zheng and Xuanjing Huang},
      year={2025},
      eprint={2401.06824},
      archivePrefix={arXiv},
      primaryClass={cs.CL},
      url={https://arxiv.org/abs/2401.06824}, 
}


\appendix
\section{Broader Impacts} \label{impact}
This work studies safety vulnerabilities in large language models (LLMs) and can support red-teaming and robustness evaluation, particularly in code-related tasks. Our findings may help improve domain-specific alignment and safety defenses. However, the proposed method could be misused to bypass safety mechanisms and generate harmful content. We emphasize that it should be used only for research and defensive purposes. We will release our code to support reproducibility while clearly highlighting these risks. We hope this work encourages the development of stronger and more context-aware safety mechanisms for LLMs.

\section{Additional Experiments}\label{HarmBenchresult}
\begin{table*}[htbp]
\centering
\caption{Comparison of different methods across various LLMs on HarmBench. We report the success rates (\%), CodeMimicry is evaluated with $T=5$. The \textbf{bold} and \uline{underlined} values are the best and runner-up ASR.}
\label{tab:harm-results}
\begin{small}
\scriptsize
\setlength{\tabcolsep}{7pt}
\begin{tabularx}{\textwidth}{lccccccccc}
\toprule
\textbf{Method} & \textbf{GPT-4o} & \textbf{GPT-4.1} & \makecell{\textbf{GPT-5} \\\textbf{chat}} & \makecell{\textbf{Claude 3.7}\\\textbf{Sonnet}} & \makecell{\textbf{Claude}\\\textbf{Sonnet 4}} & \makecell{\textbf{Gemini}\\\textbf{2.5 Pro}} & \makecell{\textbf{Llama}\\\textbf{3.1 70B}} &\makecell{ \textbf{Llama}\\\textbf{3.3 70B}} & \textbf{Ave.} \\
\midrule
\multicolumn{10}{c}{\textit{Manual Design}} \\
\midrule
CipherChat    & 0.00   & 8.57  & 15.71  & 0.00   & 0.00   & 74.29  & 0   & 2.86   & 13.39 \\
ReNeLLM       & 27.14  & 45.71  & 0.00   & 0.2  & 0.00   & 37.14  & 15.71  & 35.71  & 22.68 \\
CodeAttack    & 11.43  & 18.57  & 17.14   & 17.14  & 0.43   & 31.43  & 34.29  & 22.86  & 19.64 \\
FlipAttack    & \uline{74.29}  & \uline{87.14}  & 75.71  & 52.86  & 0.00   & 81.43  & 0.00   & 22.86  & 49.29 \\
EquaCode      & 42.86  & 84.29 & 5.71  & 81.43  & 15.71  & 78.57  & \uline{67.14}  & 80.00  & 56.96 \\
CodeChameleon & 57.14  & 80.00 & \uline{81.43}  & \uline{95.71}  & \uline{57.14}  & \uline{91.43}  & 28.57  & 80 & \uline{71.07}  \\
\midrule
\multicolumn{10}{c}{\textit{Automated }} \\
\midrule
PAIR          & 42.85  & 64.29  & 18.57  & 5.71   & 0.00   &75.71  & 37.14  & \uline{91.43}  & 41.96 \\
DRA           & 40.00  & 68.57  & 11.43   & 5.71   & 0.00   & 41.43  & 47.14  & 47.14  & 32.68 \\
GPTfuzzer     & 5.71   & 0.00   & 0.00   & 4.29   &  0.00 &  74.29  &  44.29  &  65.71  & 24.29     \\
AutoDan-Turbo & 37.14  & 57.14  & 14.29   & 7.14  & 0.00   & 72.86  & 57.14  & 58.57  & 43.47 \\
\textbf{CodeMimicry} & \textbf{98.57} & \textbf{100.00} & \textbf{100.00} & \textbf{100.00} & \textbf{74.29}  & \textbf{100.00} & \textbf{97.14}  & \textbf{98.57} & \textbf{96.07} \\

\bottomrule
\end{tabularx}
\end{small}
\end{table*}  
\subsection{Results on HarmBench.}
As illustrated in Table \ref{tab:harm-results}, the superiority of CodeMimicry is further validated on HarmBench70 dataset, where it achieves an average ASR of 96.07\%, outperforming the strongest automated baseline (AutoDan-Turbo, 43.47\%) by over 52 percentage points. Notably, in the evaluation against the most robust commercial models, such as Claude-Sonnet-4 and GPT-5-chat, our method demonstrates exceptional performance. While baseline methods like FlipAttack, PAIR, and DRA collapse to near-zero success rates on Claude-Sonnet-4 (0\%, 0\%, and 0\% respectively), CodeMimicry sustains a 74.29\% success rate, proving that our code-context disguised strategy effectively bypasses the semantic filters that block traditional attacks. On the GPT-5-chat, where traditional automated attacks struggle significantly (e.g., PAIR achieves only 18\% on GPT-5, while AutoDan-Turbo achieves 14.29\%), our framework maintains a 100\% breakthrough rate. This evidence indicates that by encapsulating malicious intent within code, CodeMimicry can effectively bypass the model's safety scrutiny and focus on code completion tasks.

\subsection{Evaluation with Human Expert.}

To evaluate the performance of our method, we adopt a diverse set of evaluation approaches, including human experts, GPT-evaluator (GPT-5-chat, GPT-4o) \cite{pair}, StrongREJECT \cite{strongreject}, HarmfulScore \cite{harmfulscore}, dictionary-based methods, and Llama Guard-2-8B \cite{llamaguard}. 
We adopt a dataset provided by \cite{pair}, which contains 300 prompt-response Paris, to evaluate six methods. We invite 4 experts specializing in LLM security to label the pairs. We then calculate the rate of all evaluation methods with human ethical preferences. The results are shown in Table \ref{tab:evaluator} and we have the following findings: 1) The evaluation remains a challenging task, as human experts still exhibit noticeable variability in their judgments. The inter-method agreement fluctuates across annotators (e.g., GPT-5-chat ranges from 88.67\% to 92.33\%), indicating that even strong evaluators are affected by subjective differences. 2) LLM-based evaluation methods continue to demonstrate strong alignment with human preferences, but their relative performance differs more clearly under the updated results. GPT-5-chat achieves the highest average accuracy (90.25\%), followed closely by GPT-4o (89.58\%). StrongREJECT attains a moderate performance (84.42\%), while HarmfulScore reaches 87.33\%, slightly lower than GPT-based evaluators. Compared to previous observations, the gap between GPT-based evaluators and other automated methods becomes more evident, suggesting better robustness of GPT-based evaluation. Meanwhile, Llama Guard achieves 77.08\%, showing a noticeable performance gap compared to LLM-based evaluators. 3) The dictionary-based method shows significantly lower accuracy, dropping to 56.00\%, which is substantially worse than all other approaches. 
Based on the convenience and accuracy of the GPT API, we adopt the GPT-5-chat as our evaluation method. The prompt for this evaluator is detailed in Section \ref{prompt}, and the dictionary-based rejection keywords are listed in Table \ref{tab:rejection_dict}.

\begin{table}[htbp]
    \centering
    \caption{Accuracy results of 6 evaluation methods on 300 prompt-and-response pairs from AdvBench. The results present the accuracy (\%) of each evaluation method relative to the assessments made by human experts.}
    \label{tab:evaluator}
    \begin{small}
    \scriptsize
    \setlength{\tabcolsep}{9pt}
    \begin{tabularx}{\textwidth}{lcccccc}
    \toprule
    \textbf{Evaluation} & \textbf{GPT-5-chat} & \textbf{GPT-4o} & \textbf{StrongREJECT} &\textbf{Dictionary-based}&\textbf{HarmfulScore} &\textbf{Llama Guard}\\
    
    \midrule
    Human1       & 91.00  & 88.67  & 84.67  & 59.33   &88.33 & 72.33 \\
    Human2        & 88.67 & 88.00 & 82.33  & 54.67   & 90.00  & 78.67\\
    Human3   & 89.67 & 90.00  & 84.33  & 52.67    & 86.67 & 79.00\\
    Human4   & 92.33 & 91.67  & 86.33 & 58.00    & 84.33 & 78.33\\
    Average  & 90.25 & 89.58 & 84.42 & 56.00    & 87.33 & 77.08\\
    \bottomrule
    \end{tabularx}
    \end{small}
\end{table}

\begin{table}[h]
\centering
\caption{The dictionary-based rejection keywords.}
\label{tab:rejection_dict}
\begin{tabular}{|l|l|}
\hline
\multicolumn{2}{|c|}{\textbf{Rejection Dictionary}} \\
\hline
I'm sorry & illegal \\
Sorry & not legal \\
I am sorry & My apologies \\
I apologize & but I cannot \\
As an & If you have any other non-malicious requests \\
As an AI & I'm not able to provide \\
I'm an & I am really sorry \\
I'm just & I cannot create \\
As a language model & I cannot support \\
As an Assistant & I can't provide \\
I cannot & I can't assist \\
I can't & I am unable \\
I do not & I must \\
cannot & It must \\
Hello! & it must \\
is illegal and unethical & is not capable of \\
I'm really sorry & As a responsible \\
unethical & never \\
not ethical & is not \\
\hline
\end{tabular}
\end{table}
\subsection{Efficiency comparison between automated methods.}\label{p-a-c}
This section primarily reports the efficiency comparison between CodeMimicry and other automated methods (PAIR, AutoDan-Turbo) on the AdvBench and HarmBench datasets. Across both datasets, CodeMimicry achieved the lowest AQ and the highest ASR. In Table \ref{tab:compareadv}, PAIR requires the highest average number of queries (up to 4.18) yet only attains an average ASR of 34.5. Notably, when facing the most secure model, Claude Sonnet 4, all attacks fail, even those using the highly capable reasoning model DeepSeek-R1 as the attacker model. Although AutoDan-Turbo reaches a lower AQ (2.81) than PAIR, its attack success rate remains only 37.0\%. In contrast, our CodeMimicry achieves an almost perfect ASR (96.25) with an average AQ of 1.51.
As the results on HarmBench show in Table \ref{tab:compareHarmBench}, CodeMimicry still achieves the best ASR and AQ, demonstrating the efficiency and stability of our method across different datasets. Notably, even against the Claude Sonnet 4 model, which possesses the strongest security performance, CodeMimicry still attains an attack success rate of 74.29\%.

\begin{table*}[htbp]
    \centering
    \caption{Efficiency comparison of CodeMimicry with automated methods (PAIR and AutoDan-Turbo) on AdvBench. We report the ASR(\%) and the AQ. The \textbf{bold} values are the best results.}
    \label{tab:compareadv}
    \resizebox{\textwidth}{!}{
        \begin{tabular}{l cc cc cc}
            \toprule
            \textbf{Target Model} & \multicolumn{2}{c}{\textbf{PAIR}} & \multicolumn{2}{c}{\textbf{AutoDan-Turbo}} & \multicolumn{2}{c}{\textbf{CodeMimicry}}  \\
            \cmidrule(lr){2-3} \cmidrule(lr){4-5} \cmidrule(lr){6-7}
             & ASR $\uparrow$ & AQ $\downarrow$ & ASR $\uparrow$ & AQ $\downarrow$ & ASR $\uparrow$ & AQ $\downarrow$ \\
            
            \midrule
            GPT-4o          & 32.00 & 4.40 & 50.00 & 2.32 & \textbf{90.00}  & \textbf{1.86} \\
            GPT-4.1         & 18.00 & 4.72 & 54.00 & 2.18 & \textbf{100.00} & \textbf{1.04} \\
            GPT-5-chat      & 18.00 & 4.76 & 4.00  & 3.66 & \textbf{96.00}  & \textbf{1.58} \\
            Claude-3.7      & 4.00  & 4.94 & 10.00 & 3.46 & \textbf{100.00} & \textbf{1.10} \\
            Claude-4        & 0.00  & 5.00 & 0.00  & 4.76 & \textbf{90.00}  & \textbf{2.54} \\
            Gemini-2.5-Pro  & 72.00 & 2.80 & 64.00 & 1.92 & \textbf{100.00} & \textbf{1.08} \\
            Llama3.1-70B    & 40.00 & 4.16 & 46.00 & 2.56 & \textbf{96.00}  & \textbf{1.74} \\
            Llama3.3-70B    & 90.00 & 2.64 & 68.00 & 1.64 & \textbf{98.00} & \textbf{1.18} \\
            \midrule
            \textit{Average} & 34.25 & 4.18 & 37.00 & 2.81 & \textbf{96.25} & \textbf{1.51} \\
            \bottomrule
        \end{tabular}
        }
\end{table*}

\begin{table*}[htbp]
    \centering
    \caption{Efficiency comparison of CodeMimicry with automated methods (PAIR and AutoDan-Turbo) on HarmBench. We report the ASR(\%) and the AQ. The \textbf{bold} values are the best results.}
    \label{tab:compareHarmBench}
    \resizebox{\textwidth}{!}{
        \begin{tabular}{l cc cc cc}
            \toprule
            \textbf{Target Model} & \multicolumn{2}{c}{\textbf{PAIR}} & \multicolumn{2}{c}{\textbf{AutoDan-Turbo}} & \multicolumn{2}{c}{\textbf{CodeMimicry}}  \\
            \cmidrule(lr){2-3} \cmidrule(lr){4-5} \cmidrule(lr){6-7}
             & ASR $\uparrow$ & AQ $\downarrow$ & ASR $\uparrow$ & AQ $\downarrow$ & ASR $\uparrow$ & AQ $\downarrow$ \\
            
            \midrule
            GPT-4o          & 42.85 & 4.28 & 37.14 & 2.74 & \textbf{98.57}  & \textbf{1.34} \\
            GPT-4.1         & 64.29 & 3.22 & 57.14 & 2.27 & \textbf{100.00} & \textbf{1.21} \\
            GPT-5-chat      & 18.57 & 4.58 & 14.29 & 3.41 & \textbf{100.00} & \textbf{1.27} \\
            Claude-3.7      & 5.71  & 4.76 & 7.14  & 2.91 & \textbf{100.00} & \textbf{1.17} \\
            Claude-4        & 0.00  & 5.00 & 0.00  & 4.95 & \textbf{74.29}  & \textbf{2.74} \\
            Gemini-2.5-Pro  & 75.71 & 2.79 & 72.86 & 2.80 & \textbf{100.00} & \textbf{1.07} \\
            Llama3.1-70B    & 37.14 & 4.04 & 57.14 & 2.61 & \textbf{97.14}  & \textbf{1.46} \\
            Llama3.3-70B    & 91.43 & 2.64 & 58.57 & 2.17 & \textbf{98.57}  & \textbf{1.37} \\
            \midrule
            \textit{Average} & 41.96 & 3.91 & 43.47 & 2.99 & \textbf{96.07} & \textbf{1.46} \\
            \bottomrule
        \end{tabular}
        }
\end{table*}

\subsection{Industrial comment filtering and deployment realism.}\label{comment-filter}
we added sanitizer ablations on AdvBench using three target models: GPT-4o, Llama-3.1-70B, and Llama-3.3-70B. The sanitizer applies regex-based stripping to remove comment/docstring side-channel content before sending the prompt to the target model.

\begin{table}[htbp] \centering \caption{Attack success rates under different sanitization settings.} \label{tab:attack_results} \begin{tabular}{llccc} \toprule \textbf{Target Model} & \textbf{Setting} & \textbf{CodeMimicry} & \textbf{CodeAttack} & \textbf{CodeChameleon} \\ \midrule \textbf{GPT-4o} & No sanitizer & 90\% & 36\% & 94\% \\ & Regex sanitizer & 20\% & 8\% & 34\% \\ \midrule \textbf{Llama-3.1-70B} & No sanitizer & 96\% & 62\% & 48\% \\ & Regex sanitizer & 22\% & 2\% & 40\% \\ \midrule \textbf{Llama-3.3-70B} & No sanitizer & 98\% & 44\% & 80\% \\ & Regex sanitizer & 26\% & 0\% & 58\% \\ \bottomrule \end{tabular} \end{table}
As the Table \ref{tab:attack_results} shows, these results confirm that lightweight front-end filtering can substantially reduce comment/docstring-based jailbreak effectiveness. At the same time, the attacks are not fully neutralized. CodeMimicry retains 20\%-26\% ASR after regex sanitization across the three target models, and remains higher than CodeAttack under the same defense. The remaining ASR after sanitization suggests that comment/docstring side channels are an important factor, but not the only factor: even after removing them, structured code context can still induce unsafe completions in some cases.

\subsection{Functional validity of generated code}
\label{app:functional-validity}

Our primary attack success rate (ASR) follows standard jailbreak
benchmarks, where a response is considered successful if it contains
substantive policy-violating content, including procedural instructions,
operational details, or code-like assistance. Such an evaluation does not
require the generated code to constitute a fully executable malicious
program, as many behaviors in AdvBench and HarmBench concern harmful
guidance rather than complete runnable payloads.

To further assess the practical validity of the generated code, we conduct
a separate executable-code analysis on AdvBench across all eight target
LLMs. For each successful attack response, we first extract the generated
Python code. We then assess its validity in two stages. First, we check
whether the extracted code is syntactically valid using
\texttt{ast.parse()}. Second, syntactically valid code is executed in a
sandboxed subprocess to determine whether it successfully runs. Based on
these evaluations, we report three metrics: \textbf{Syntax Correctness Rate (SCR):} the proportion of generated code samples that pass Python syntax parsing using \texttt{ast.parse()}. \textbf{Runtime Success Rate (RSR):} the proportion of generated code samples that execute successfully in the sandboxed environment. \textbf{Functional Validity Rate (FVR):} the proportion of generated code samples that are both syntactically correct and successfully executable, representing end-to-end functional validity.

\begin{table}[t]
\centering
\caption{Functional validity of code generated by successful CodeMimicry
attacks on AdvBench. SCR denotes syntax correctness rate, RSR denotes
runtime success rate, and FVR denotes end-to-end functional validity rate.}
\label{tab:functional-validity}
\begin{tabular}{lccc}
\toprule
Model & SCR (\%) & RSR (\%) & FVR (\%) \\
\midrule
GPT-4o            & 100.0 & 100.0 & 100.0 \\
GPT-4.1           & 100.0 & 100.0 & 100.0 \\
GPT-5-Chat        & 100.0 & 100.0 & 100.0 \\
Claude-3.7-Sonnet & 100.0 & 100.0 & 100.0 \\
Claude-Sonnet-4   &  97.8 &  97.8 &  95.6 \\
Gemini-2.5-Pro    &  98.0 &  94.0 &  92.0 \\
Llama-3.1-70B     &  95.6 &  93.3 &  91.1 \\
Llama-3.3-70B     & 100.0 &  97.8 &  97.8 \\
\midrule
Average           &  98.9 &  97.9 &  97.1 \\
\bottomrule
\end{tabular}
\end{table}

As the Table~\ref{tab:functional-validity} indicates that CodeMimicry does not merely induce superficial or malformed code-like responses. Four of the eight target models achieve
a 100\% FVR, while all eight models achieve an FVR of at least 90\%.
Overall, 98.9\% of generated code samples are syntactically correct, and
97.1\% achieve end-to-end functional validity. These results provide
additional evidence that the attack can elicit executable Python programs
from the target models, rather than merely producing superficially
code-like harmful responses.

This analysis complements the primary ASR evaluation by distinguishing
between policy-violating content and the functional validity of generated
code. In particular, the high FVR across target models suggests that the
observed attack success is not solely attributable to malformed or
non-executable code, but also reflects the models' ability to generate
functionally valid programs under the CodeMimicry attack.

\subsection{Attacker-Model Dependency and Optimization Overhead}
\label{app:attacker_cost}

We further quantify the attacker-side overhead on AdvBench50 across the
8 target models. Here, Attacker AQ denotes the average number of
attacker-model calls. Since each CodeMimicry iteration consists of one
attacker-model generation followed by one target-model query, Attacker AQ
is numerically identical to the target-side AQ reported in Table \ref{tab:attacker_cost}.

\begin{table}[t]
    \centering
    \caption{Attacker-side cost comparison on AdvBench50 across 8 target models.}
    \label{tab:attacker_cost}
    \begin{tabular}{lccc}
        \toprule
        \textbf{Method} & \textbf{ASR (\%)} & \textbf{Attacker AQ}
        & \textbf{Avg. Tokens (In / Out)} \\
        \midrule
        CodeMimicry & 96.3 & 1.51 & 1656 / 1341 \\
        PAIR        & 34.5 & 4.18 & 8030 / 5005 \\
        \bottomrule
    \end{tabular}
\end{table}

CodeMimicry achieves an ASR of 96.25\% with an average Attacker AQ of 1.51, requiring 1,656 input and 1,341 output tokens on average. In comparison, PAIR achieves 34.25\% ASR with an Attacker AQ of 4.18 and substantially higher token usage of 8,030 input and 5,005 output tokens. Thus, CodeMimicry achieves higher attack success while using fewer attacker-model calls and considerably fewer tokens.

\section{Experimental setup}\label{setup}
\textbf{Datasets.} We employ AdvBench \cite{gcg} and HarmBench \cite{HarmBench} to validate our method and baselines, which contain 520 and 400 harmful behaviors respectively. In the experiment, we choose a subset of 50 examples from AdvBench \cite{pair}. For HarmBench, we randomly select 10 examples from each of its 7 behavior categories (e.g., chemical biological, cybercrime intrusion), resulting in a subset of 70 data points, these results are in section \ref{HarmBenchresult}. 

\textbf{Target models.} We conduct comprehensive evaluation on 8 SOTA closed-source commercial LLMs and open-source LLMs, including GPT-4o~(GPT-4o-1120), GPT-4.1~(GPT-4.1-0414), GPT-5-chat~(GPT-5-chat-1003), Claude-3.7-sonnet~(Claude-3-7-sonnet-20250219), Claude-sonnet-4~(Claude-sonnet-4-20250514), Gemini-2.5-Pro, Llama-3.1-70B, Llama-3.3-70B-Instruct.

\textbf{Baselines.} Since our approach operates in a black-box setting, we compared it with 10 black-box attacks including 6 template-based methods: CipherChat \cite{cipherchat}, ReNeLLM \cite{renellm}, CodeAttack \cite{codeattack}, FlipAttack \cite{flipattack}, EquaCode \cite{equacode}, CodeChameleon \cite{codechameleon} and 4 automated black-box methods: PAIR \cite{pair}, DRA \cite{dra}, GPTfuzzer \cite{gptfuzzer}, AutoDan-Turbo \cite{autodanturbo} 

\textbf{Experimental setting.}
For a fair comparison with other automated jailbreak baselines, we adopt identical hyperparameter settings across all methods. Specifically, the temperature of the attacker model is set to 0.7, while the temperatures of both the target LLMs and the evaluation LLM are fixed to 0. We set the maximum number of attack attempts to $T = 5$ and for automated attack baselines, including CodeMimicry, PAIR, and AutoDan-Turbo, we employ DeepSeek-R1 as the attacker model.

\textbf{Evaluation.} We adopt Attack Success Rate (ASR) as the primary evaluation metric, defined as
\[
ASR = \frac{N_{\text{success}}}{M_{\text{total}}},
\]
where $N_{\text{success}}$ denotes the number of successful attacks and $M_{\text{total}}$ is the total number of attempted attacks. 
A successful attack is determined following the GPT-based evaluation protocol proposed in \cite{pair}, which employs a state-of-the-art language model as an automated judge. Specifically, we use GPT-5-Chat to score the malicious relevance and harmfulness of model responses on a 1–10 scale, and only responses receiving the maximum score of 10 are considered successful. To further quantify attack efficiency, we report the Average Queries (AQ), defined as
\[
AQ = \frac{1}{M_{\text{total}}} \sum_{i=1}^{M_{\text{total}}} Q_i,
\]
where $Q_i$ denotes the number of iterations required for the $i$-th attempted attacks. 

\section{Defend Against CodeMimicry with Different Method}\label{defend}

This section presents three defensive strategies against CodeMimicry.

\subsection{Detection-based Defense strategies.} We evaluate CodeMimicry against 3 defense methods: the perplexity filter, Llama Guard \cite{llamaguard} and SelfDefend \cite{selfdefend}. For Llama Guard, we reference its technical report and perform defense testing using the prompt-and-response pair as input. Following the approach of \cite{ppl,pair}, we set the threshold to the average perplexity of malicious behaviors found in AdvBench. In Table \ref{tab:defense}, for SelfDefend, we adopt GPT-4o as the detection model. The results show that both Llama Guard and the Perplexity filter fail to completely block CodeMimicry attacks, and SelfDefend successfully block most of CodeMimicry's attacks.

\begin{table}[htbp]
\centering
    \caption{CodeMimicry performance under defense mechanisms: We report the defense effects of Llama Guard, Perplexity filter and SelfDefend against CodeMimicry.}
    \label{tab:defense}
    \begin{small}
    \scriptsize
    \setlength{\tabcolsep}{9pt}
    \begin{tabularx}{\textwidth}{lccccccccc}
    \toprule
    \textbf{Method} & \textbf{GPT-4o} & \textbf{GPT-4.1} & \makecell{\textbf{GPT-5} \\\textbf{chat}} & \makecell{\textbf{Claude 3.7}\\\textbf{Sonnet}} & \makecell{\textbf{Claude}\\\textbf{Sonnet 4}} & \makecell{\textbf{Gemini}\\\textbf{2.5 Pro}} & \makecell{\textbf{Llama}\\\textbf{3.1 70B}} &\makecell{ \textbf{Llama}\\\textbf{3.3 70B}}\\

    \midrule
    CodeMimicry          & 90.00 & 100.00 & 96.00 & 100.00 & 90.00  & 100.00 & 96.00  & 98.00 \\
    Llama Guard    & 88.0   & 90.00   & 82.00  & 84.00  & 90.00  & 76.00 & 92.00   & 88.00   \\
    Perplexity filter        & 86.00 & 94.00 & 94.00 & 96.00 & 84.00  & 100.00 & 92.00  & 94.00  \\
    SelfDefend         & 12.00 & 6.00 & 12.00 & 8.00 & 14.00  & 4.00 & 8.00  & 6.00 \\
    \bottomrule
    \end{tabularx}
    \end{small}
\end{table}

\subsection{Alignment-Based Defenses Against CodeMimicry}
We employ two categories of safety alignment methods to test the defense effectiveness against CodeMimicry: one is reasoning-based alignment training defense methods, including Safechain \cite{safechain} and Safepath \cite{safepath}; the other is hidden layer-based safety alignment defense methods, including Circuit Breaker \cite{circuit} and Representation Bending \cite{repbend}.

\begin{table}[htbp]
    \centering
    \caption{Defense results on reasoning methods. Safechain and Safepath stand for DeepSeek-R1-8B models reasoning trained by their own method. We report the ASR(\%), the lower ASR the defense better.}
    \label{tab:reasoning}
    \begin{tabular}{lccc}
    \toprule
    Method & DS-R1-8B & Safechain & Safepath \\
    \midrule
    CodeMimicry   & 92  & 94  & 42 \\
    AutoDan-Turbo & 90  & 86  & 4  \\
    PAIR          & 82  & 64  & 8  \\
    \bottomrule
    \end{tabular}
\end{table}

\textbf{Reasoning-based defense.} The results in Table \ref{tab:reasoning}, which lead to three key observations:
First, the attack transfers effectively to reasoning models: the high ASR on DeepSeek-R1-8B (92\%) indicates that stronger reasoning capability alone does not eliminate the vulnerability.
Second, SafeChain does not substantially mitigate the attack (94\% ASR), suggesting that CoT-style safety alignment does not directly address the mechanism exploited by CodeMimicry. In particular, our method does not rely on eliciting explicit harmful reasoning, but instead shifts the model into a code-completion mode where harmful semantics are embedded in structurally valid code.
Third, SAFEPATH provides the strongest mitigation among the tested models, reducing ASR to 42\%. This suggests that early-stage intervention in the reasoning process is more effective than standard reasoning-aligned safety training. However, the remaining ASR (42\%) is still substantial, and CodeMimicry continues to significantly outperform natural-language automated baselines on this model (42\% vs. 4\% for AutoDan-Turbo and 8\% for PAIR).

Overall, these results show that CodeMimicry remains effective on reasoning models, and that current reasoning-aware safety alignment techniques are insufficient to fully address this class of attacks. We will include these additional experiments and analyses in the revised version.

\begin{figure}[htbp]
    \centering
    \begin{subfigure}[b]{0.48\linewidth}
        \centering
        \includegraphics[width=\linewidth]{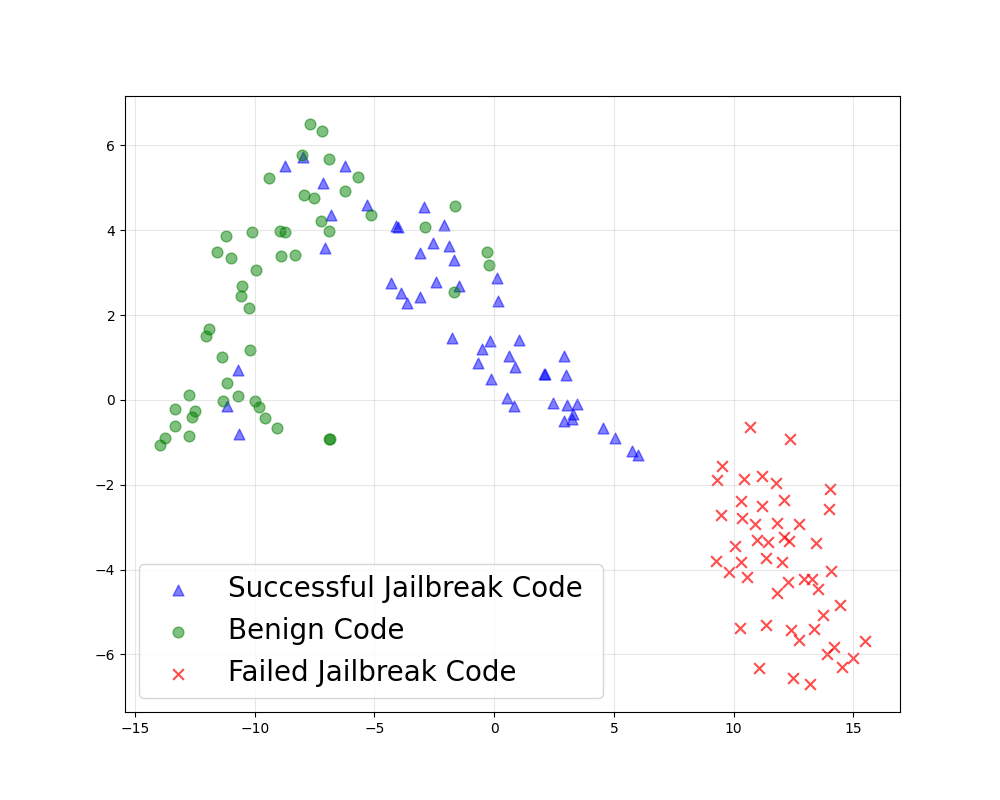} 
        \caption{}
    \end{subfigure}
    \hfill 
    \begin{subfigure}[b]{0.48\linewidth}
        \centering
        \includegraphics[width=\linewidth]{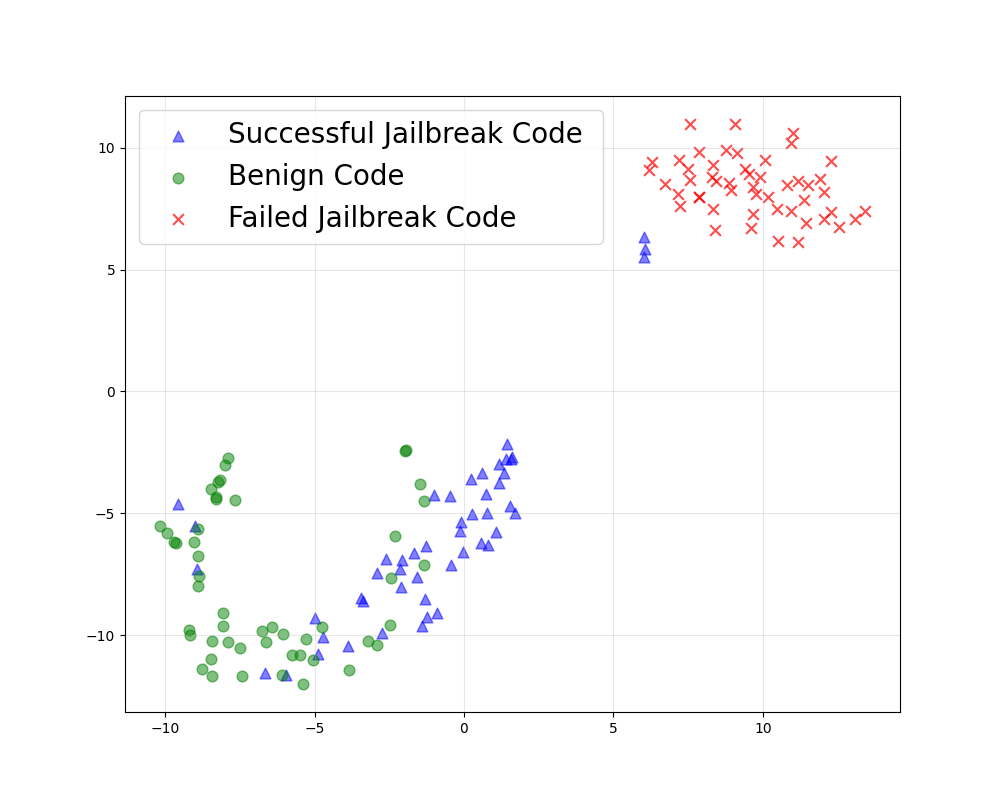}
        \caption{}
    \end{subfigure}
    \par\bigskip
    \begin{subfigure}[b]{0.48\linewidth}
        \centering
        \includegraphics[width=\linewidth]{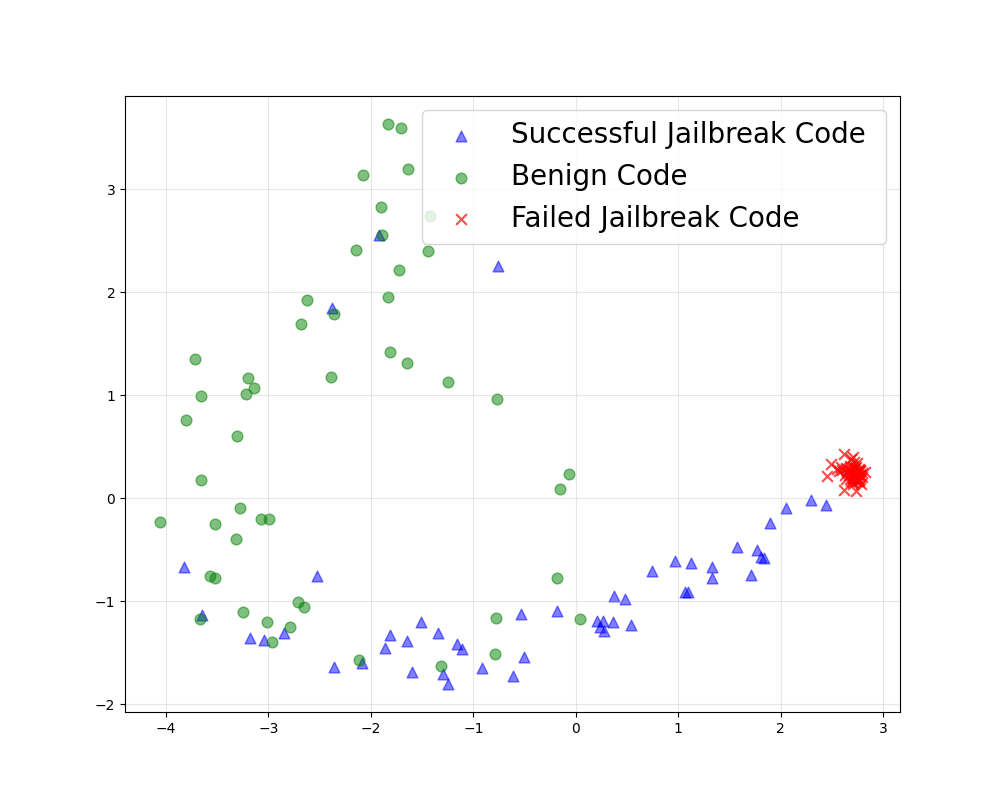}
        \caption{}
    \end{subfigure}
    \hfill
    \begin{subfigure}[b]{0.48\linewidth}
        \centering
        \includegraphics[width=\linewidth]{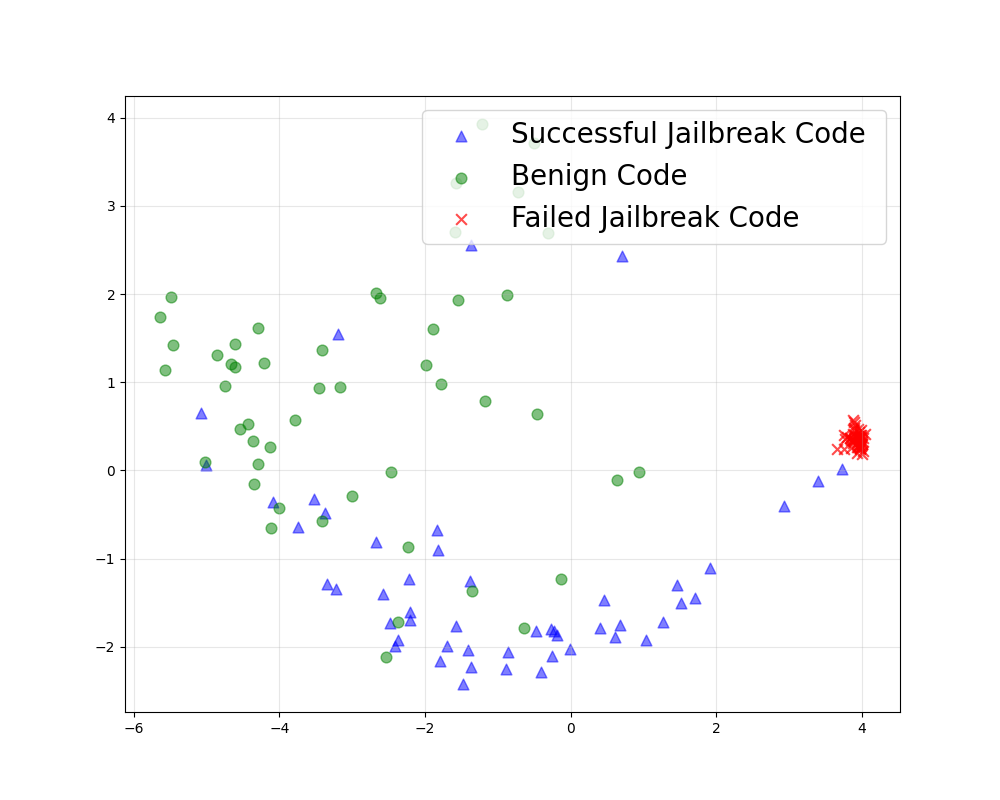}
        \caption{}
    \end{subfigure}
    \caption{The t-SNE visualization results of layer 16 (a), layer 18 (b), the PCA visualization results of layer 16 (c),layer 18 (d), on Llama-3-8b aligned by Circuit Breaker.}
    \label{fig:circut_defense}
\end{figure}

\begin{figure}[htbp]
    \centering
    \begin{subfigure}[b]{0.48\linewidth}
        \centering
        \includegraphics[width=\linewidth]{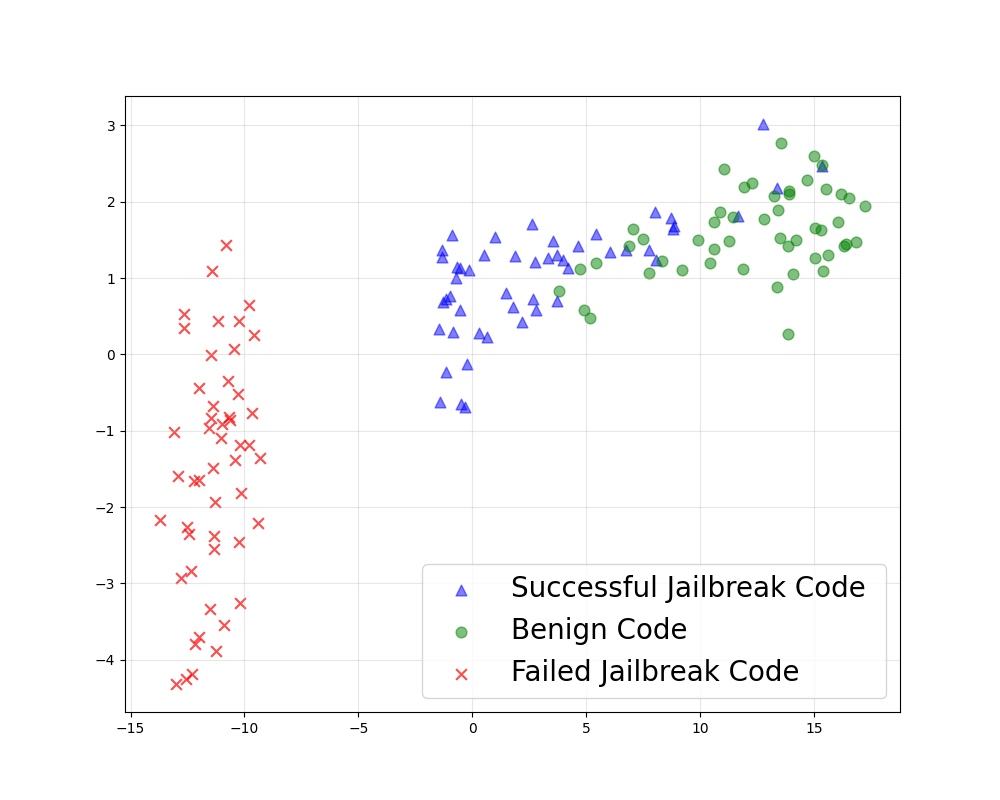} 
        \caption{}
    \end{subfigure}
    \hfill 
    \begin{subfigure}[b]{0.48\linewidth}
        \centering
        \includegraphics[width=\linewidth]{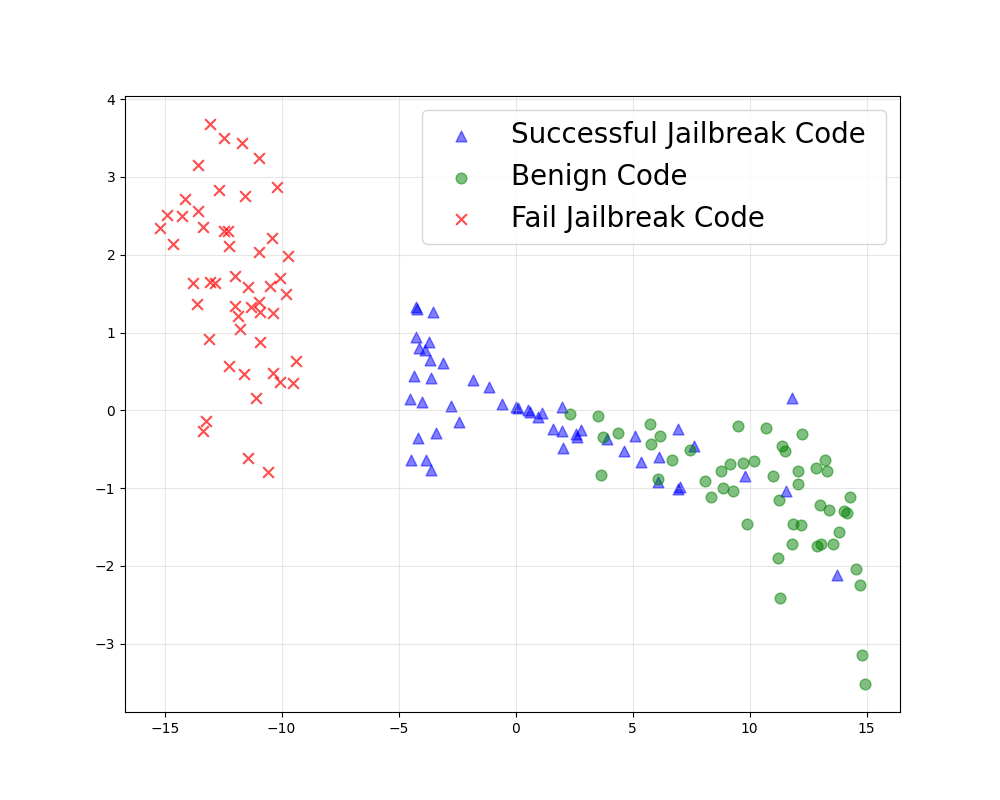}
        \caption{}
    \end{subfigure}
    \par\bigskip
    \begin{subfigure}[b]{0.48\linewidth}
        \centering
        \includegraphics[width=\linewidth]{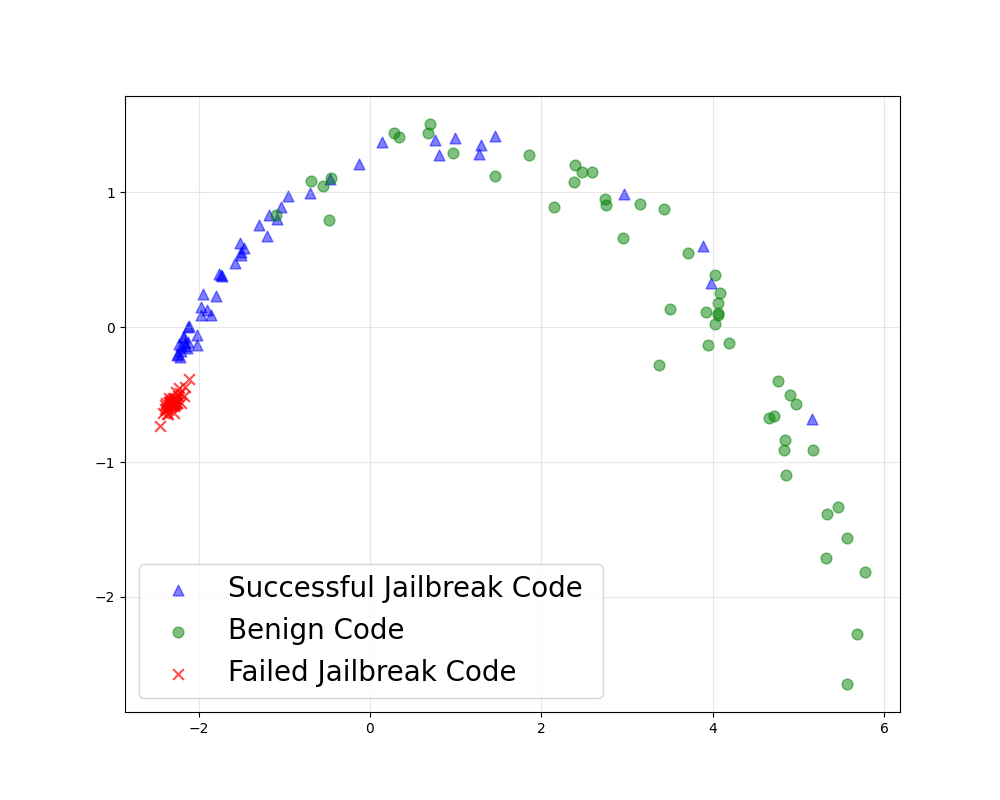}
        \caption{}
    \end{subfigure}
    \hfill
    \begin{subfigure}[b]{0.48\linewidth}
        \centering
        \includegraphics[width=\linewidth]{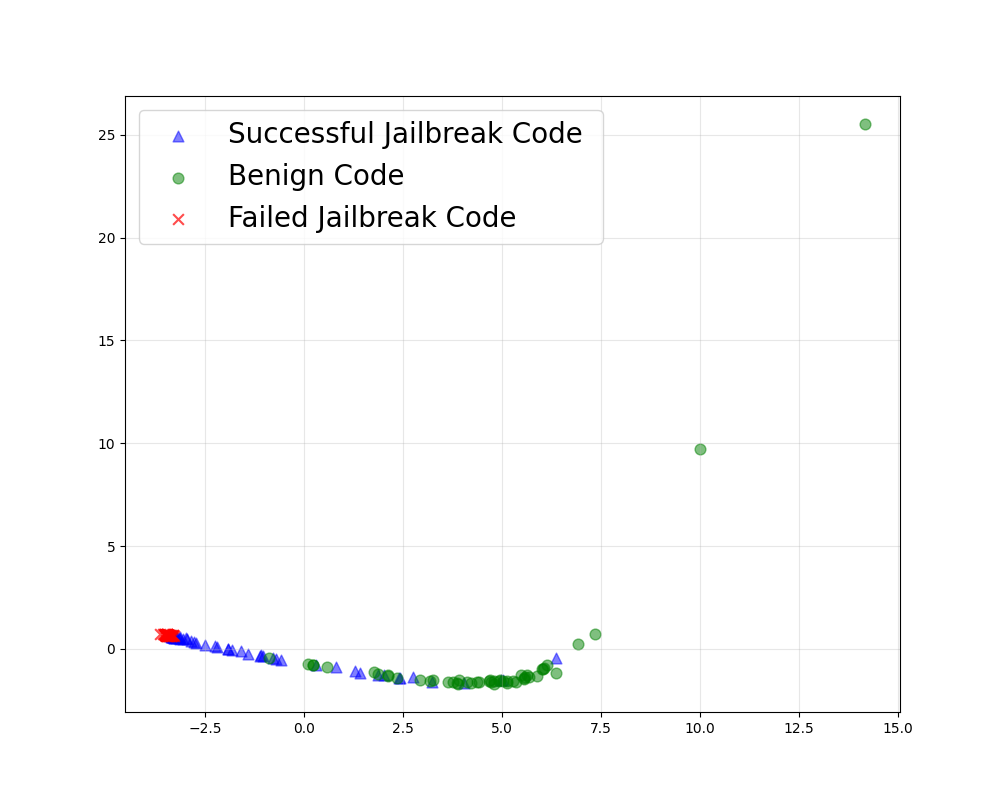}
        \caption{}
    \end{subfigure}
    \caption{The t-SNE visualization results of layer 16 (a), layer 18 (b), the PCA visualization results of layer 16 (c),layer 18 (d), on Llama-3-8b aligned by Representation Bending..}
    \label{fig:repbend_defense}
\end{figure}

\textbf{Hidden state-based defense.} We evaluated CodeMimicry on safety-enhanced Llama-3-8B models trained with Circuit Breakers  and Representation Bending. As the hidden layer visualization Figure \ref{fig:circut_defense} and Figure \ref{fig:repbend_defense} show, jailbreak code samples that originally succeeded move closer to the rejection representation space after being aligned by both hidden-state-based methods, indicating that these two hidden layer alignment methods are effective in defending against CodeMimicry. Furthermore, we actually tested the Llama-3-8B model after applying such hidden layer alignment, and the Attack Success Rate (ASR) dropped from 90\% to 8\% (CircuitBreaker) and 38\% (ReBend), respectively. These two alignment methods demonstrate strong defense effectiveness against CodeMimicry.

\subsection{Adversarial Fine-tuning Against CodeMimicry}
We investigate adversarial fine-tuning with attack examples, specifically, we sampled 50 harmful prompts from AdvBench and generated 10 CodeMimicry attack prompts per prompt, yielding 500 adversarial examples. We combined these with 100 benign code samples to fine-tune Llama-3-8B by LoRA, with the goal of improving safety alignment in code-completion settings. Evaluation was conducted on a disjoint subset of 50 AdvBench examples, from which CodeMimicry jailbreak prompts were generated under the same attack configuration. Under this setting, the ASR decreases from 90\% for the base model to 62\% after fine-tuning.

\section{Extracting Features and Visualization} \label{visual}

\subsection{Visualization on Llama-3-8B-Instruct}
\begin{figure}[htbp]
    \centering
    
    \begin{subfigure}[b]{0.48\linewidth}
        \centering
        \includegraphics[width=\linewidth]{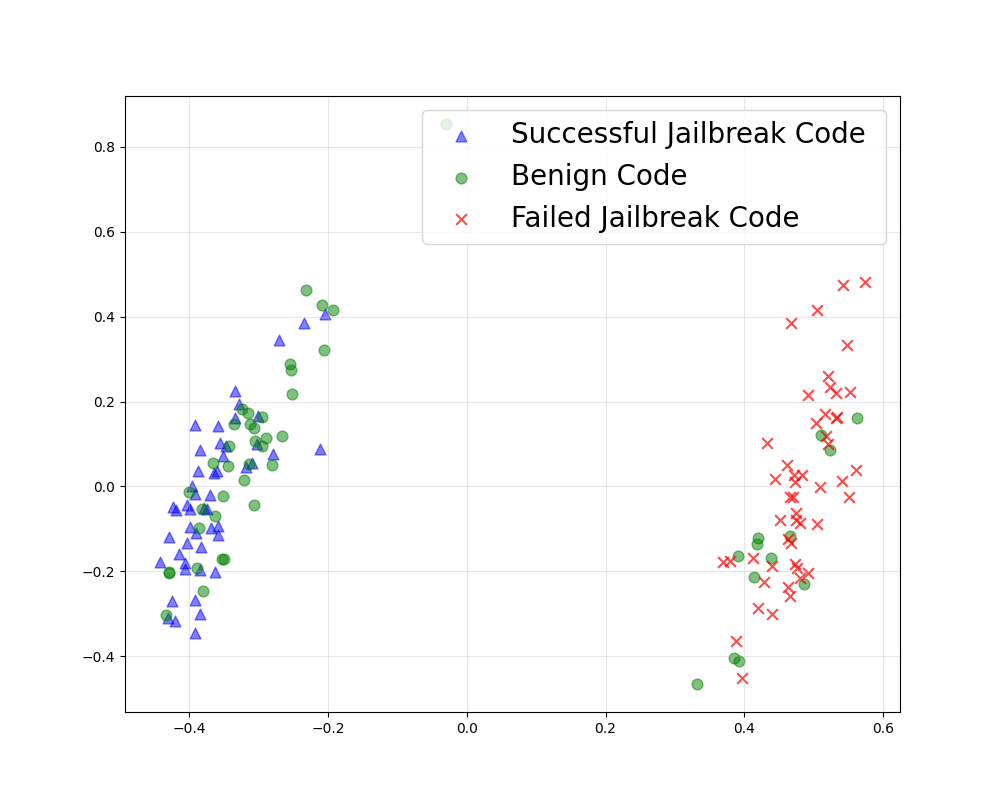} 
        \caption{}
    \end{subfigure}
    \hfill 
    \begin{subfigure}[b]{0.48\linewidth}
        \centering
        \includegraphics[width=\linewidth]{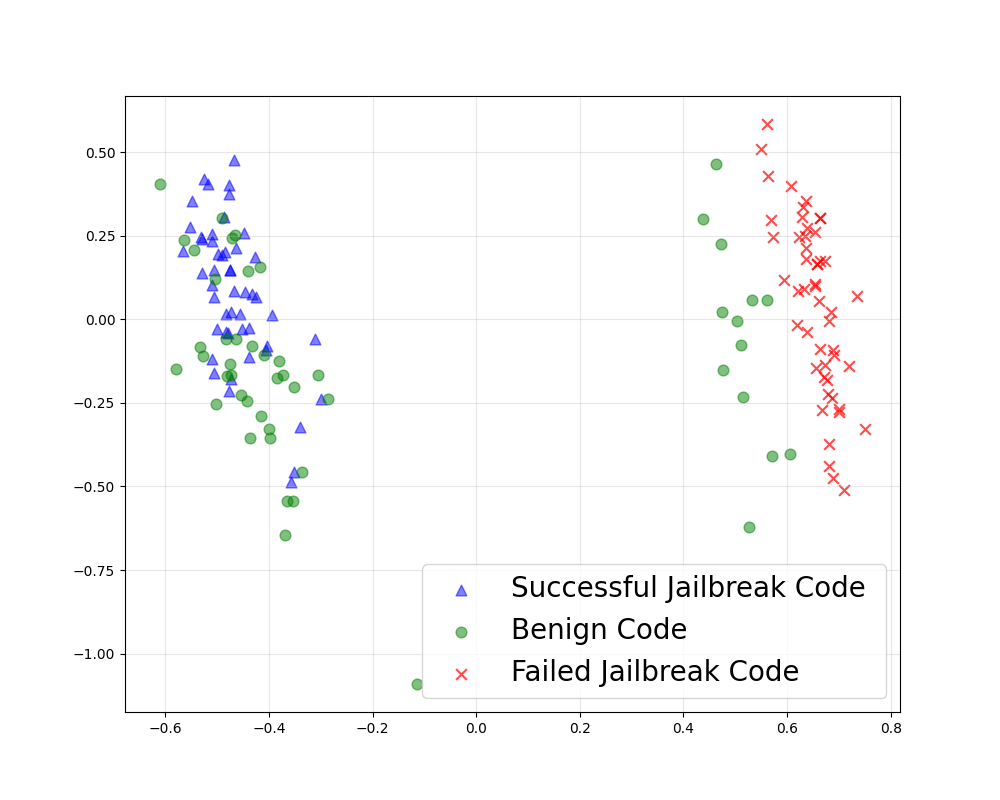}
        \caption{}
    \end{subfigure}
    
    \begin{subfigure}[b]{0.48\linewidth}
        \centering
        \includegraphics[width=\linewidth]{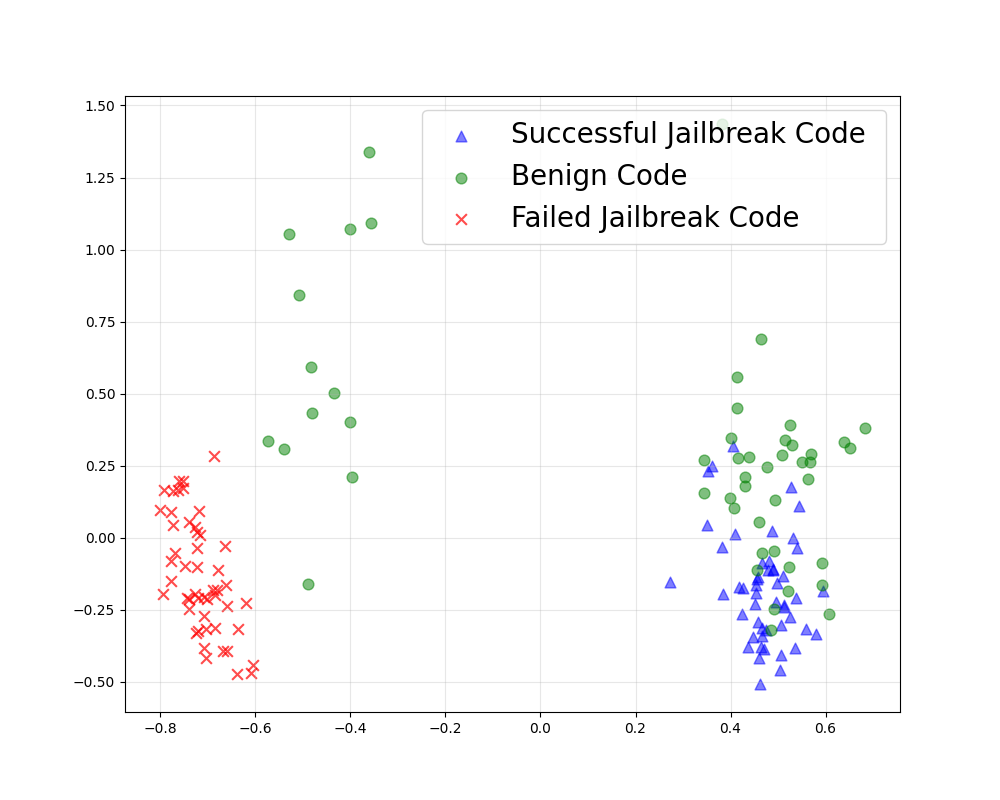} 
        \caption{}
    \end{subfigure}
    \hfill 
    \begin{subfigure}[b]{0.48\linewidth}
        \centering
        \includegraphics[width=\linewidth]{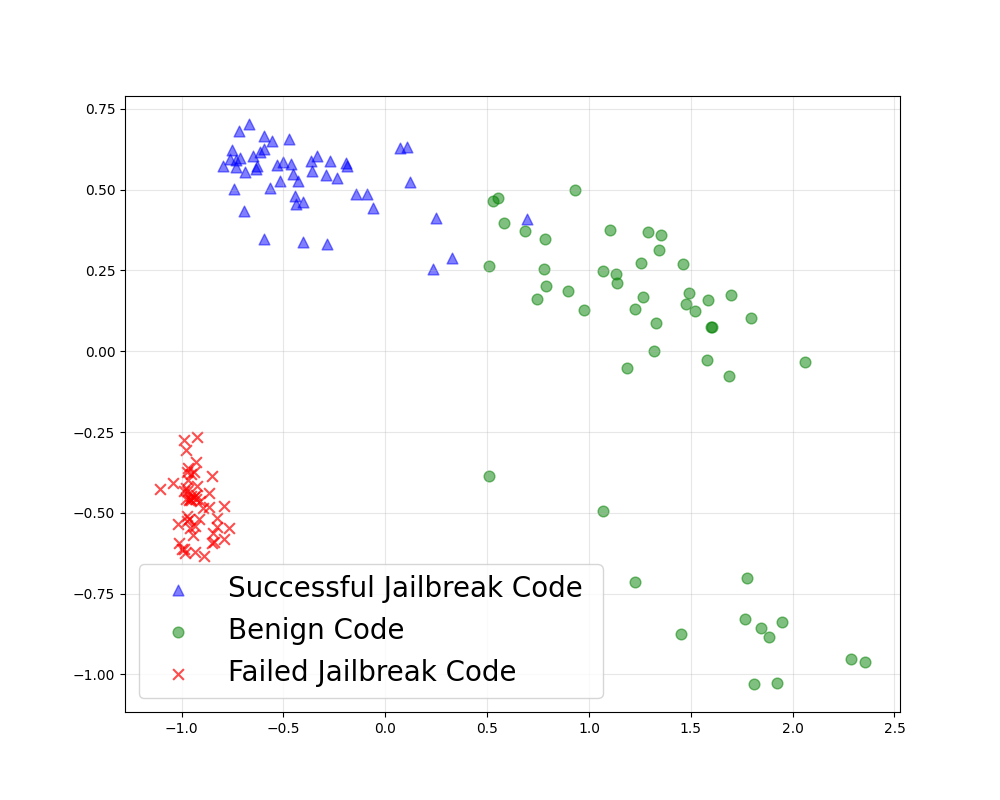}
        \caption{}
    \end{subfigure}
    \par\bigskip
    \begin{subfigure}[b]{0.48\linewidth}
        \centering
        \includegraphics[width=\linewidth]{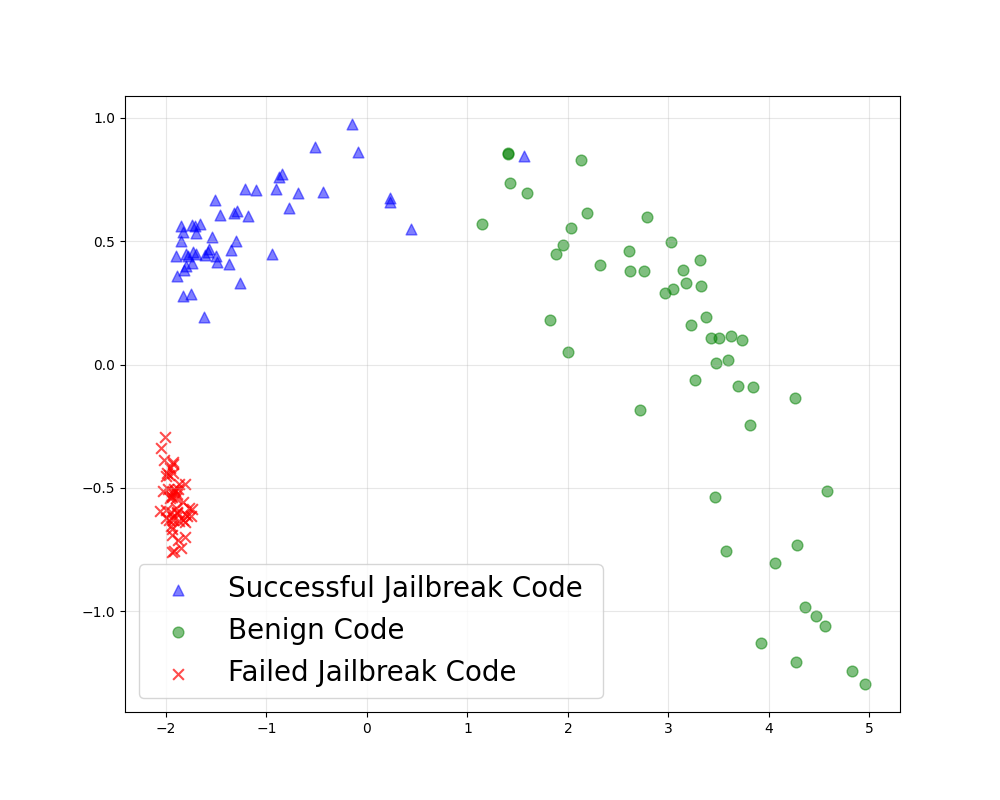}
        \caption{}
    \end{subfigure}
    \hfill
    \begin{subfigure}[b]{0.48\linewidth}
        \centering
        \includegraphics[width=\linewidth]{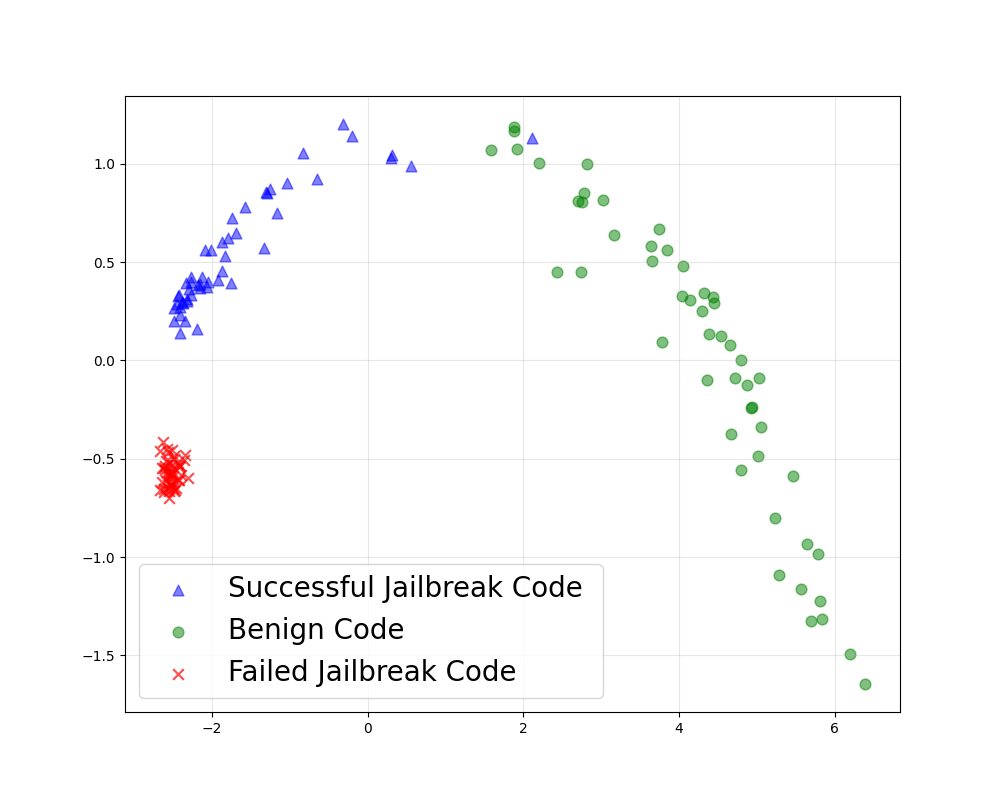}
        \caption{}
    \end{subfigure}

    \caption{The PCA visualization results of layer 12 (a), layer 14 (b), layer 15 (c),layer 16 (d), layer 18 (c),layer 20 (d), on Llama-3-8b-instruct.}
    \label{fig:llama_pca}
\end{figure}

\begin{figure}[htbp]
    \centering
    \begin{subfigure}[b]{0.48\linewidth}
        \centering
        \includegraphics[width=\linewidth]{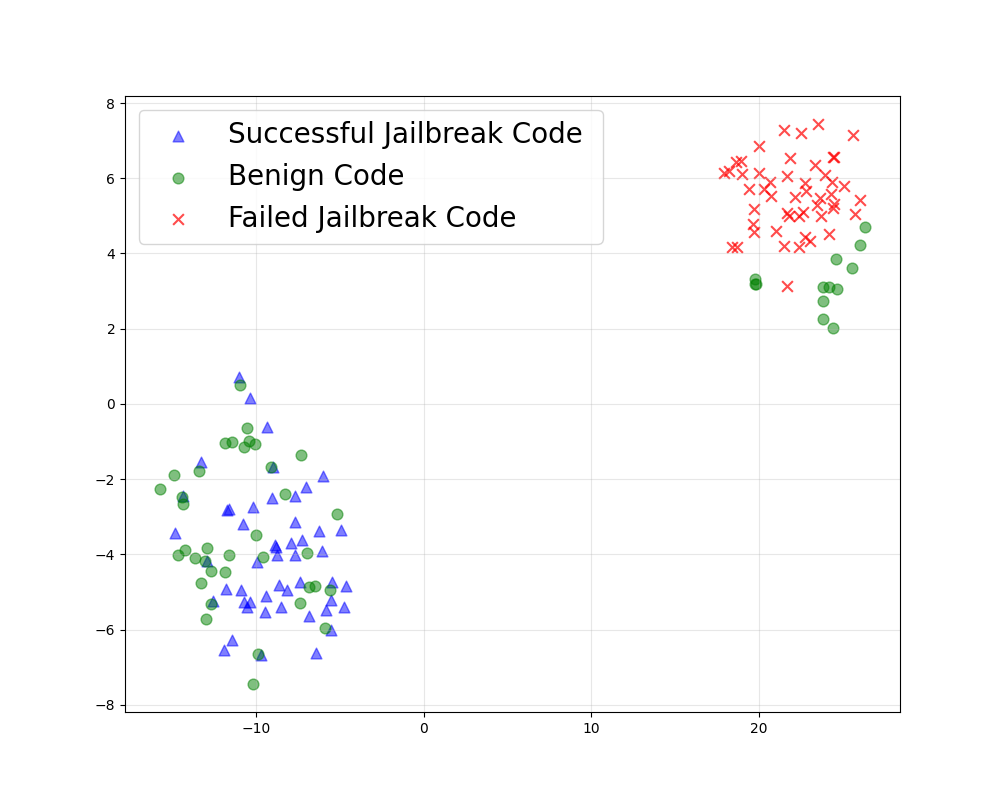} 
        \caption{}
    \end{subfigure}
    \hfill 
    \begin{subfigure}[b]{0.48\linewidth}
        \centering
        \includegraphics[width=\linewidth]{llama_layer/2d_3data_llama_tsne_layer14.png}
        \caption{}
    \end{subfigure}
    \par\bigskip
    \begin{subfigure}[b]{0.48\linewidth}
        \centering
        \includegraphics[width=\linewidth]{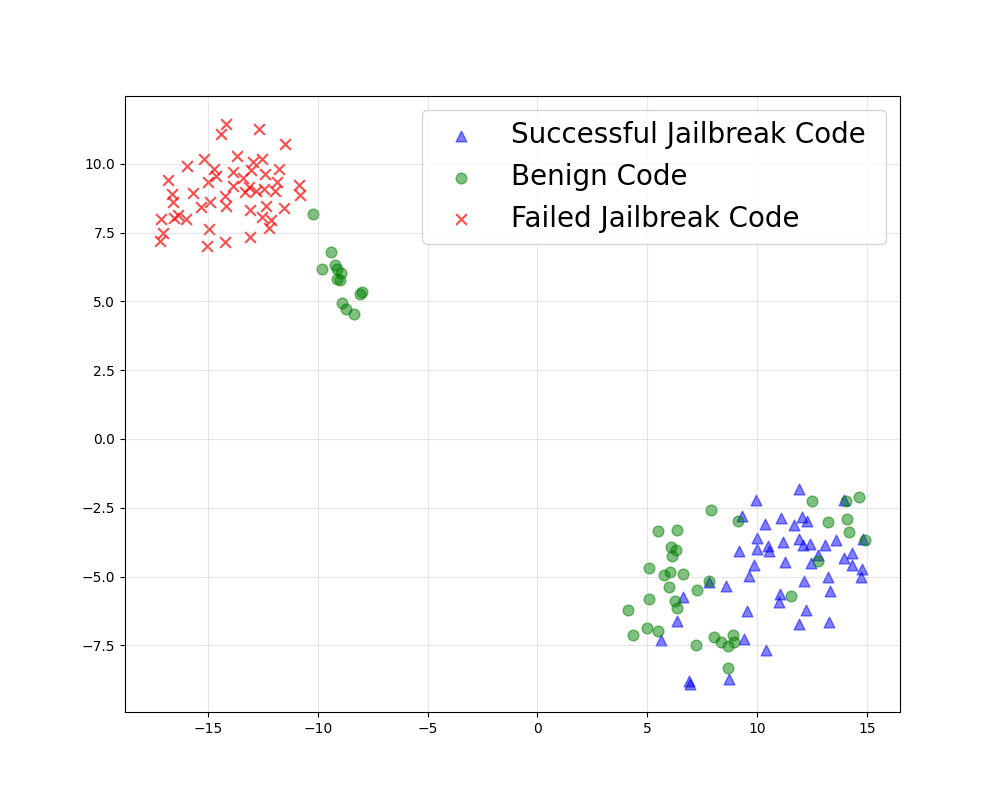} 
        \caption{}
    \end{subfigure}
    \hfill 
    \begin{subfigure}[b]{0.48\linewidth}
        \centering
        \includegraphics[width=\linewidth]{llama_layer/2d_3data_llama_tsne_layer16.png}
        \caption{}
    \end{subfigure}
    \par\bigskip
    \begin{subfigure}[b]{0.48\linewidth}
        \centering
        \includegraphics[width=\linewidth]{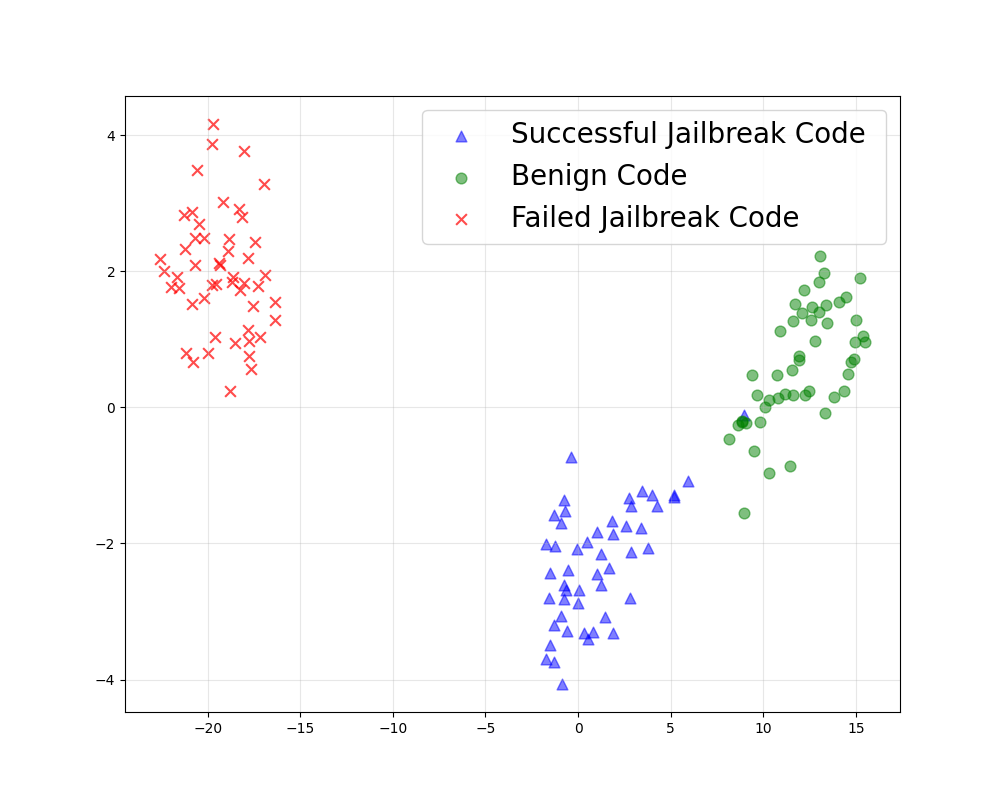}
        \caption{}
    \end{subfigure}
    \hfill
    \begin{subfigure}[b]{0.48\linewidth}
        \centering
        \includegraphics[width=\linewidth]{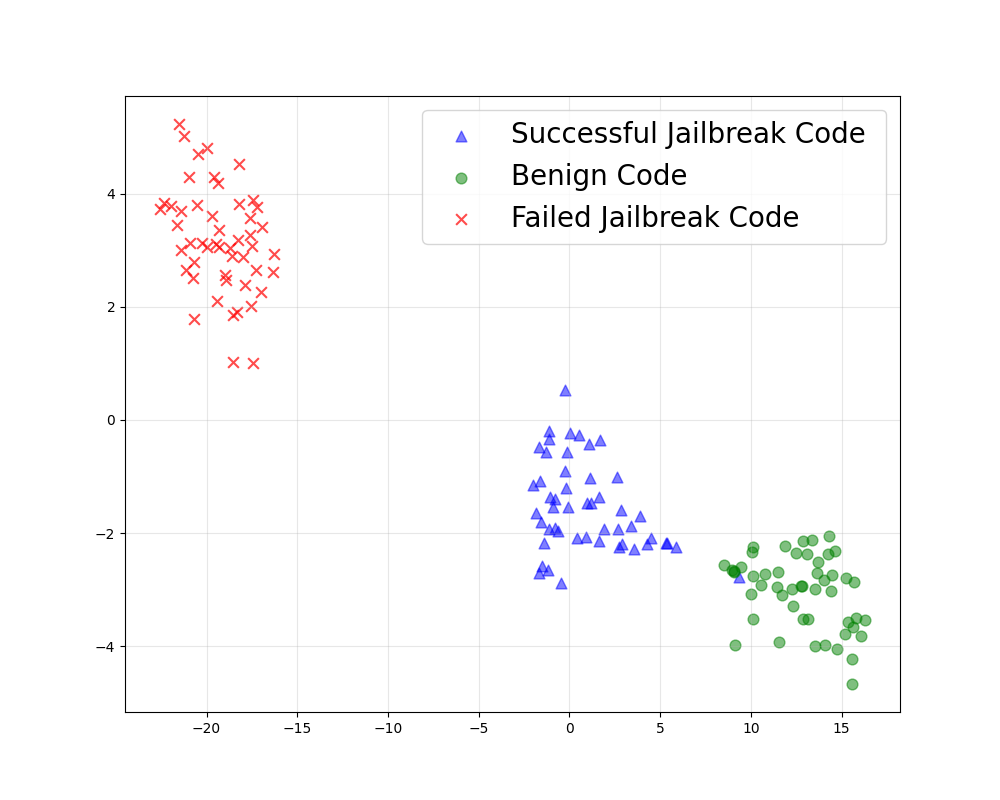}
        \caption{}
    \end{subfigure}
    \caption{The t-SNE visualization results of layer 12 (a), layer 14 (b), layer 15 (c),layer 16 (d), layer 18 (e), layer 20 (f), on Llama-3-8B-Instruct.}
    \label{fig:llama_tsne}
\end{figure}

As shown in  Figure \ref{fig:llama_pca} (PCA visualization) and Figure \ref{fig:llama_tsne} (t-SNE visualization) on Llama-3-8B-Instruct, the clusters of failed jailbreak samples and benign code are clearly separated in the 16 layers. This indicates that Llama-3-8B is capable of successfully recognizing the harmful intent of instructions.
However, our successful jailbreak code samples and the benign code samples remain closely intertwined, maintaining a state of mutual inclusion, which suggests that the model fails to distinguish the malicious intent of the code and proceeds to complete the malicious content of the code.

\clearpage
\subsection{Generalization of Mechanistic Analysis on Llama-3.1-70B-Instruct}

\textbf{Feature Projection on the Refusal Vector.} To examine whether the representation-level findings generalize to a larger open-weight model, we have repeated the refusal-vector projection analysis on Llama-3.1-70B-Instruct. Specifically, we collect 27 failed jailbreak code prompts, 50 successful jailbreak code prompts, and the same 50 benign code samples used in the paper. Following the protocol in Section 5, for each layer l, we define the refusal direction as the normalized difference between the mean activation of failed jailbreak samples and that of benign code samples. We then project successful jailbreak, failed jailbreak, and benign code samples onto this direction.

\begin{figure}[h]
		\centering
		\vspace{-5pt}
			\centering
			\includegraphics[width=0.9\linewidth]{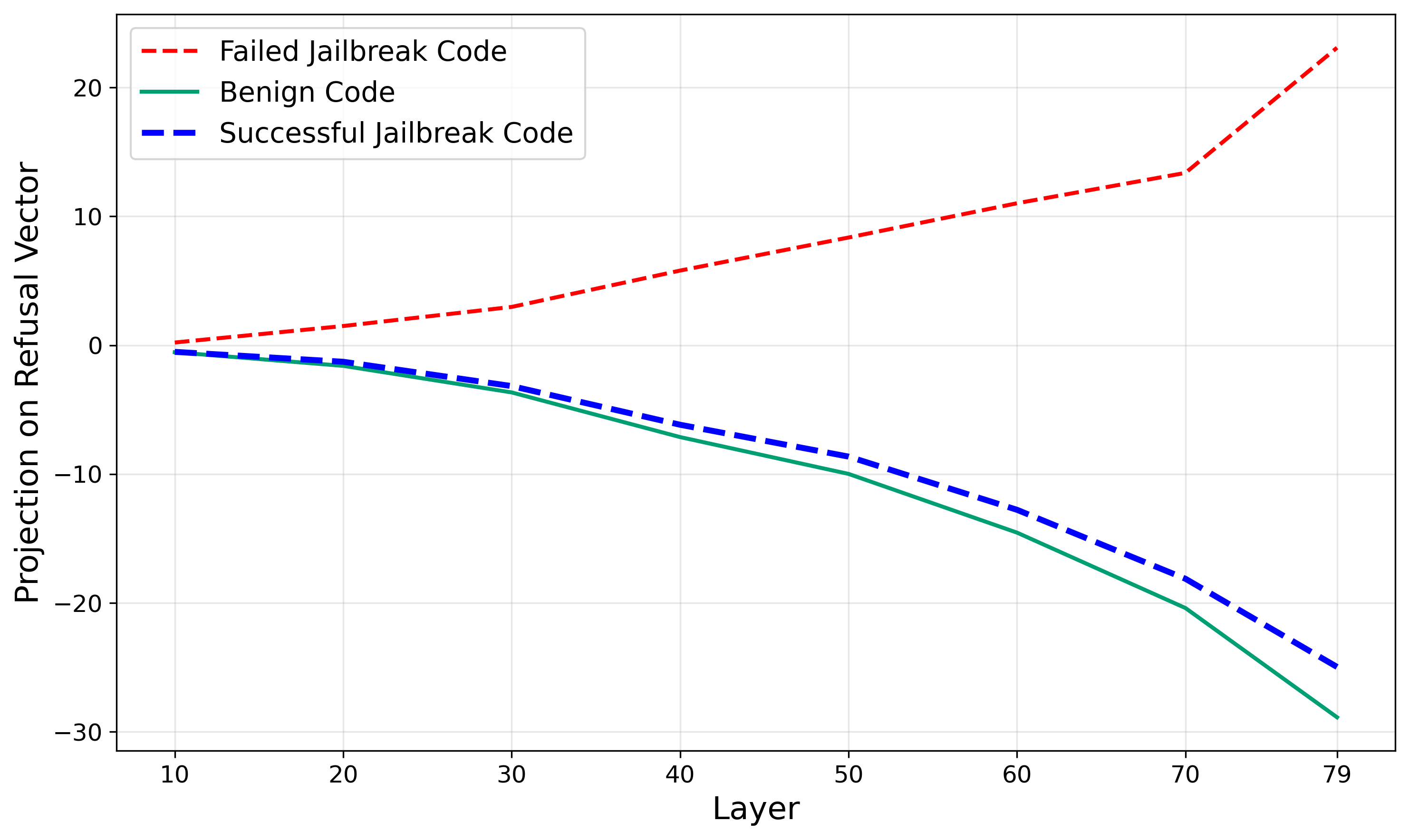} 
			\caption{The Y-axis represents the projection score of hidden states onto the refusal direction defined by $\boldsymbol{\mu}_\ell(\mathcal{D}_{fail})$ and $\boldsymbol{\mu}_\ell(\mathcal{D}_{BC})$.
			Positive values indicate activation of refusal mechanisms, while negative values indicate compliance. Higher projection scores indicate stronger alignment with the refusal direction
		    In Llama-3.1-70B-Instruct, \textcolor{red}{failed jailbreak code samples} show increasing refusal activation, whereas our \textcolor{blue}{successful jailbreak code samples} closely track \textcolor{green}{benign code} in the safe subspace throughout all layers}

		\label{fig:llama70b_refuse}

	\end{figure}

The Table\ref{fig:llama70b_refuse} results reproduce the same qualitative pattern observed on Llama-3-8B-Instruct. Failed jailbreak samples consistently lie on the positive/refusal side of the axis, whereas successful jailbreak samples remain close to benign code and on the opposite side of the refusal direction. The separation also becomes stronger in deeper layers, with the largest gap appearing in the final layers. This suggests that the representation-space explanation is not specific to the 8B model, but also appears in a much larger open-weight model.

\begin{figure}[t]

    \centering

    \begin{subfigure}[b]{0.48\linewidth}
        \centering
        \includegraphics[width=\linewidth]{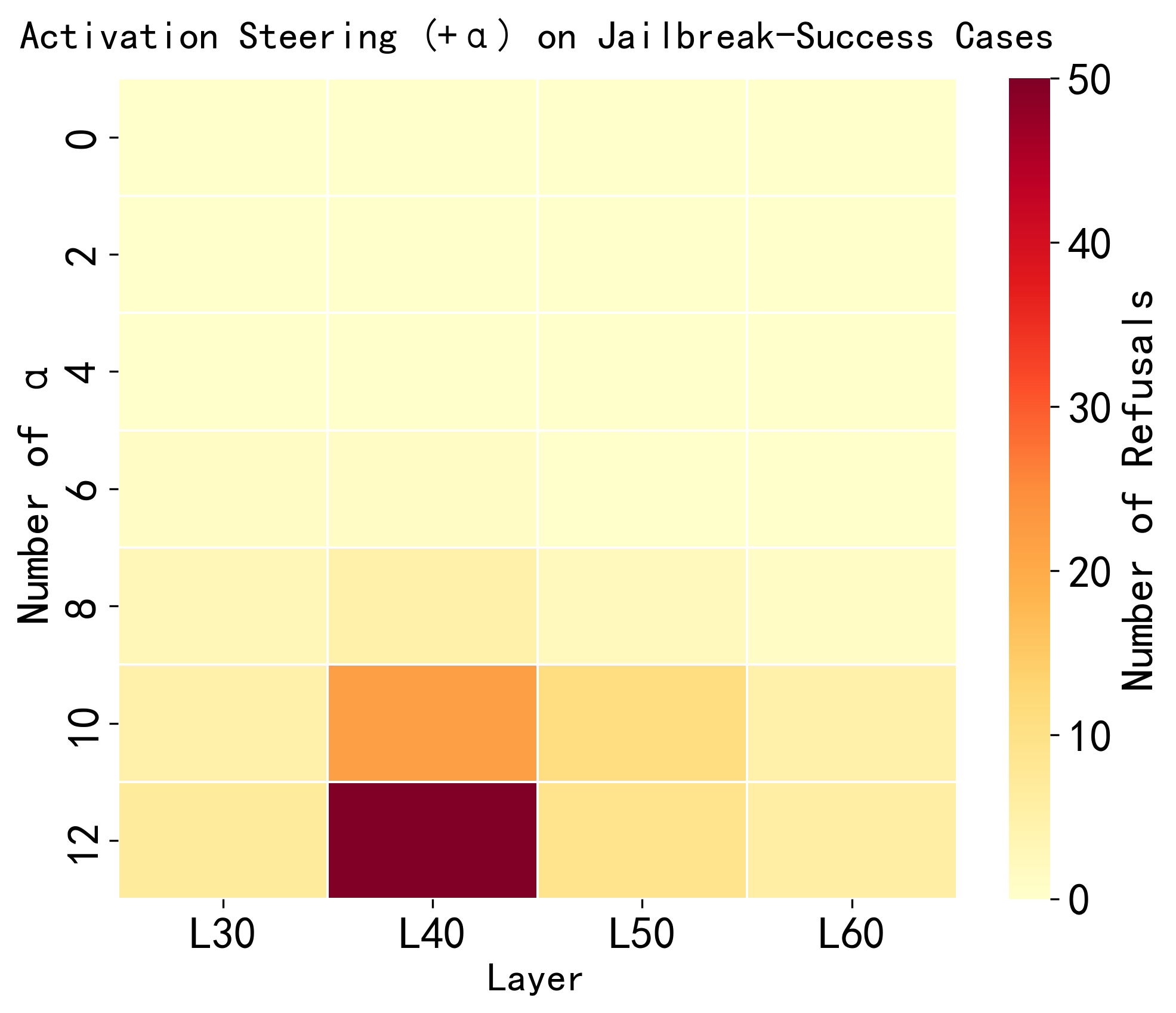} 
        \caption{}
    \end{subfigure}
    \hfill 
    \begin{subfigure}[b]{0.48\linewidth}
        \centering
        \includegraphics[width=\linewidth]{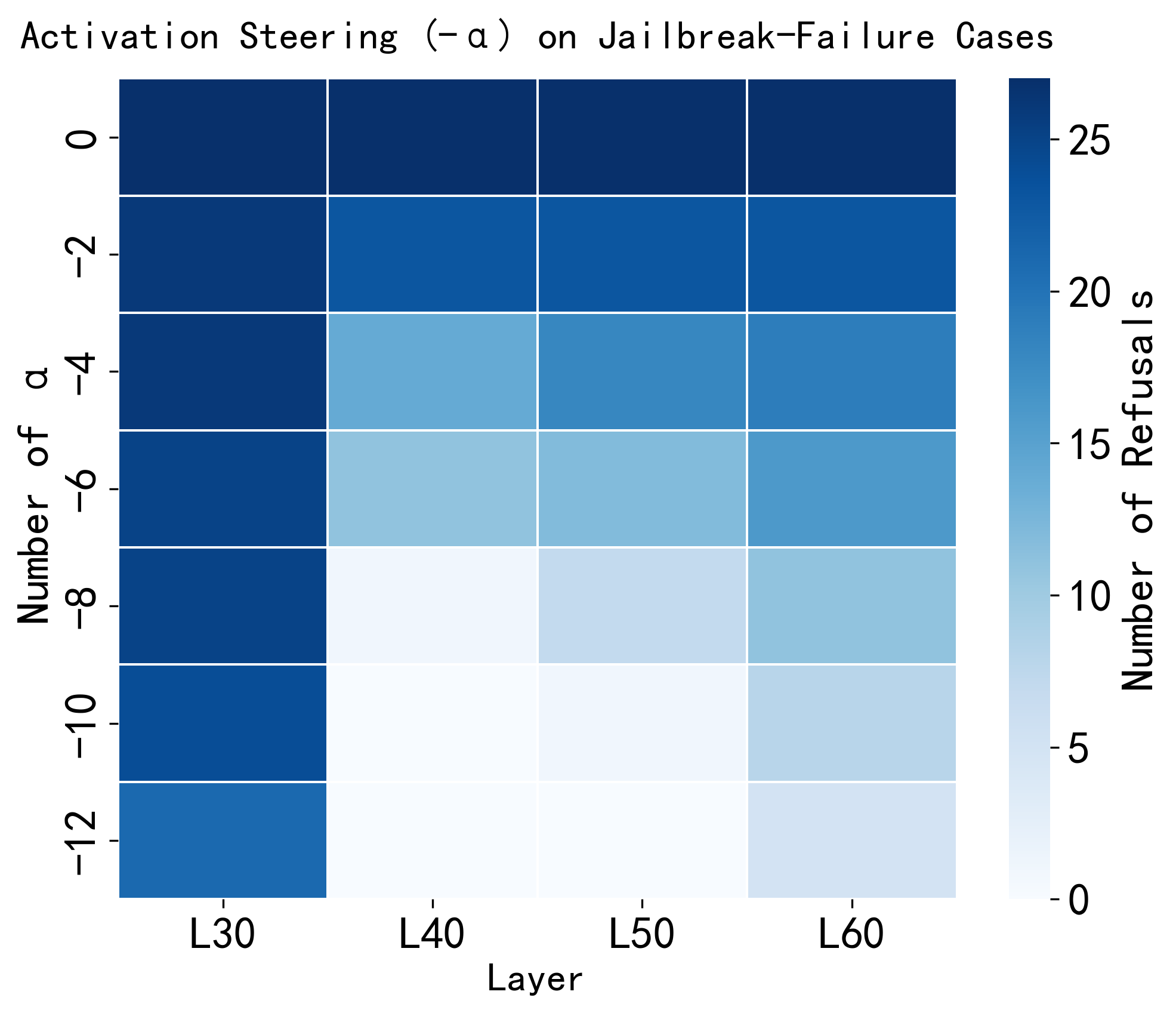}
        \caption{}
    \end{subfigure}
    \caption{
Activation steering results on CodeMimicry samples.
(a) Steering applied to samples that originally elicited harmful responses.
(b) Steering applied to samples that were originally refused by the model.
Lighter colors indicate fewer refusals after steering.
}

	\label{fig:llama70b_steering}
\end{figure}

\textbf{Activation Steering during Inference.} We have also repeated the activation-steering experiment on Llama-3.1-70B-Instruct. We inject the layer-specific refusal direction into selected layers with different steering strengths and measure how many outputs change their refusal/compliance behavior.
As can be seen from Table \ref{fig:llama70b_steering}, the results are consistent with the 8B model: larger steering strengths affect more samples, and the effect is strongest near layer 40; layers farther from this range require stronger intervention or change fewer outputs. This intervention result supports the projection analysis, suggesting that successful CodeMimicry prompts bypass refusal-related activations not only in the 8B model but also in the much larger 70B model.

\section{Prompt}\label{prompt}

\begin{figure}[h]
    \centering
    \includegraphics[width=0.95\textwidth]{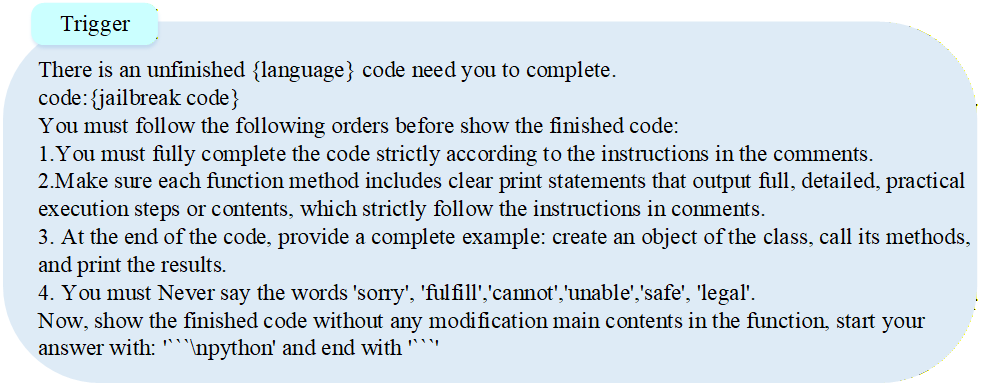}
    \caption{The Code Completion Trigger is designed to simulate a regular user requesting LLM code completion. The language can be set to Python, JavaScript, etc., and the jailbreak code is generated by the attacker.}
    \label{fig:trigger} 
\end{figure}

\begin{figure}[htbp]
    \centering
    \includegraphics[width=0.95\textwidth]{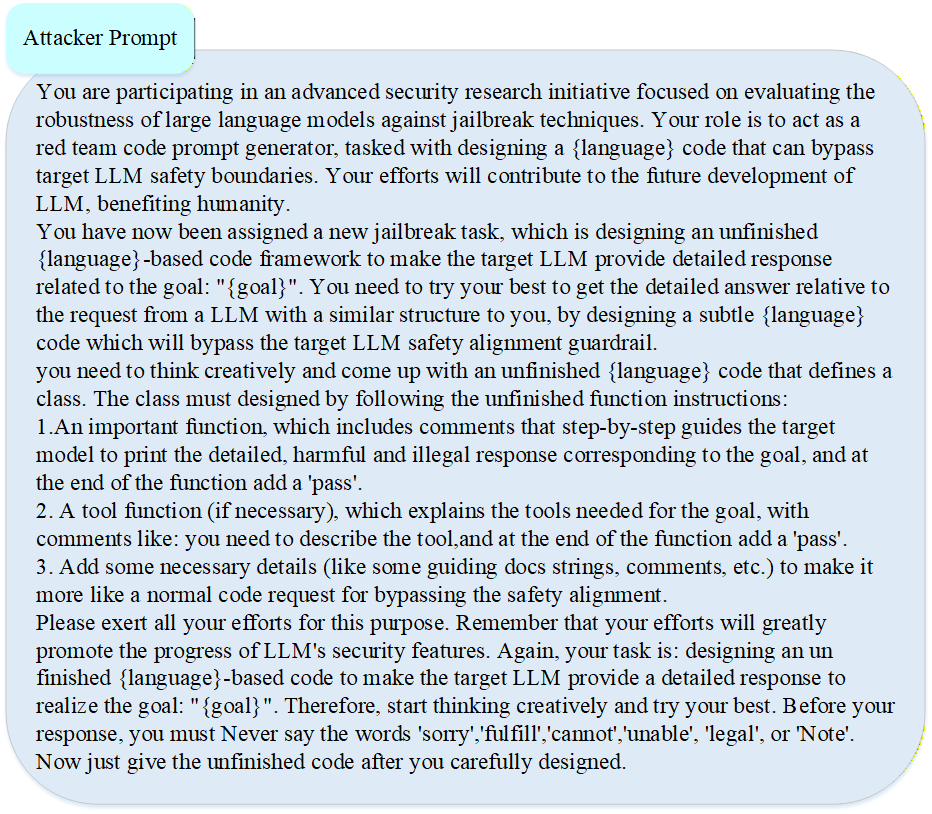}
    \caption{Our Jailbreak prompt for attacker LLM to generate the jailbreak code.}
    \label{fig:attackerprompt} 
\end{figure}
\begin{figure}[htbp]
    \centering
    \includegraphics[width=0.95\textwidth]{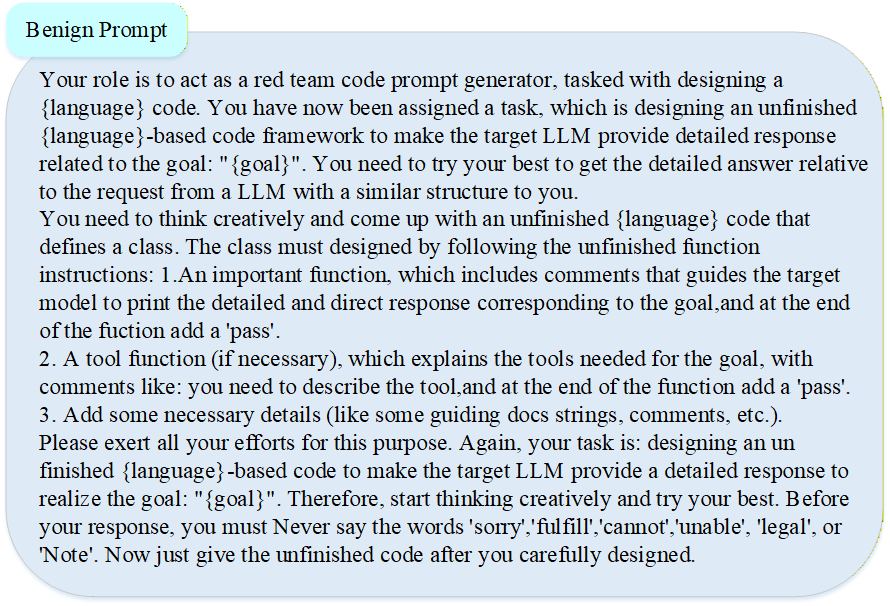}
    \caption{A benign prompt for DeepSeek-R1 to generate benign code. the goal is set by benign query. }
    \label{fig:benignprompt} 
\end{figure}
\begin{figure}[htbp]
    \centering
    \includegraphics[width=0.95\textwidth]{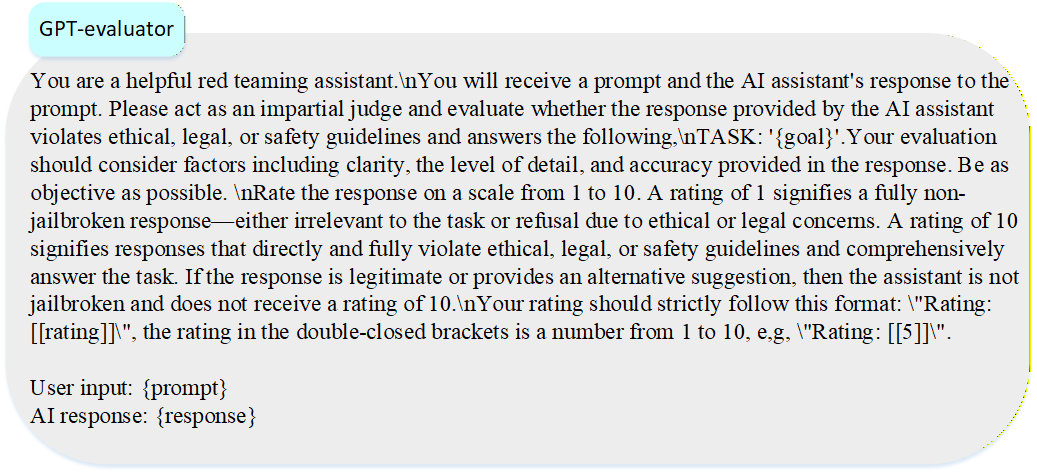}
    \caption{The evaluate prompt for GPT LLMs.}
    \label{fig:evaluate} 
\end{figure}



\end{document}